\documentclass[%
 reprint,
 amsmath,amssymb,
 aps,
nofootinbib]{revtex4-1}
\usepackage{scrextend}
\usepackage{graphicx}
\usepackage{dcolumn}
\usepackage{bm}
\usepackage{float}
\usepackage{hyperref}
\usepackage[mathlines]{lineno}
\usepackage{footnote}
\usepackage{amsmath}
\usepackage{xfrac}
\usepackage{braket}
\usepackage{cancel}
\usepackage{xcolor}
\usepackage{amssymb}

\begin{document}

\preprint{APS/123-QED}

\title{Probing Scalar and Pseudoscalar Solutions to the g-2 Anomaly}
\author{Fayez Abu-Ajamieh}
 \email{abuajamieh@ucdavis.edu, fayez.abu-ajamieh@umontpellier.fr}
\affiliation{%
Department of Physics, UC Davis, Davis, California, 95616, USA\\
LUPM UMR5299, Universit\'e de Montpellier, 34095 Montpellier, France\\
CPPM, Aix-Marseille Universit\'e, Marseille, France
}


\begin{abstract}
I investigate a class of models with scalar and pseudoscalar solutions to the $g-2$ anomaly for both the muon and the electron over the mass range of perturbativity ($m_{\phi} \lesssim 50$ GeV), with Yukawa couplings proportional to the lepton's mass. In particular, I investigate the constraints from BaBar, beam dump experiments, $Z$ decay measured quantities, LEP mono-$\gamma$ searches, $ee \rightarrow \tau\tau (\gamma)$ searches, and solar and Horizontal Branch (HB) stars bounds. For a pseudoscalar, I find that no region in the parameters space can simultaneously provide a solution for both the electron and the muon anomalies while maintaining the required form of the couplings, and therefore the pseudoscalar solution is disfavored. On the other hand, I find for the scalar case that there is an open window above $\sim 30$ MeV in the allowed region, but with significant tension with experiment for the region $m_\phi \gtrsim 10$ GeV. In addition, there is a smaller window between $\sim 350$ KeV and 1 MeV that is not ruled out by cosmological observations. Part of the first open window is expected to be covered by the proposed NA64 experiment. Similar analysis can be readily applied to other proposed solutions to the anomaly, such as solutions with $Z'$ or with the dark photon. 
\end{abstract}

\pacs{Valid PACS appear here}
\maketitle


\section{\label{sec:Introduction}Introduction}
An exciting piece of evidence for the existence of physics Beyond the Standard Model (BSM) is the discrepancy between the predicted and the measured values of the muon anomalous magnetic moment $a_{\mu} \equiv (g_{\mu}-2)/2$. The current measured value \cite{Mohr:2012tt, Bennett:2002jb, Bennett:2004pv, Bennett:2006fi} shows a $3.5\sigma$ discrepancy compared with the SM prediction \cite{Chen:2015vqy, Marciano:2016yhf, deNiverville:2018hrc}:
\begin{equation}\label{eq:muDiscrepancy}
\Delta a_{\mu} = a_{\mu}^{\text{Exp}} - a_{\mu}^{\text{SM}} = 273 \pm 80 \times 10^{-11}.
\end{equation}

A similar less significant discrepancy of about $1.1 \sigma$ was also observed for the electron \cite{Aoyama:2014sxa}:

\begin{equation}\label{eq:eDiscrepancy}
\Delta a_{e} = a_{e}^{\text{Exp}} - a_{e}^{\text{SM}} = -91 \pm 82 \times 10^{-14}.
\end{equation}

Although both discrepancies fall short of the $5\sigma$ limit required to confirm their existence, they nonetheless pose tantalizing hints for physics BSM. In addition, current experiments at Fermilab \cite{Grange:2015fou, Venanzoni:2014ixa} and at the J-PARC E34 collaboration \cite{Saito:2012zz, Mibe:2011zz} are expected to yield improved experimental results in the near future.

New physics explanations of this anomaly include (see \cite{Lindner:2016bgg} for a comprehensive review) supersymmetry (see \cite{Stockinger:2006zn} for a review), a light $Z'$ boson \cite{Langacker:2008yv, Baek:2001kca, Ma:2001md, Gninenko:2001hx, Pospelov:2008zw,Heeck:2011wj,Harigaya:2013twa, Altmannshofer:2014pba, Altmannshofer:2016brv} (also see \cite{Krasnikov:2017dmg} for a review), a scalar contribution within the framework of the 2 Higgs Doublet Model (2HDM) \cite{Iltan:2001nk, Omura:2015nja, Broggio:2014mna, Broggio:2014mna, Wang:2014sda, Abe:2015oca, Cao:2009as, Batell:2009jf, Dutta:2018hcz}, additional fermions \cite{Freitas:2014pua}, leptoquarks \cite{Chakraverty:2001yg, Cheung:2001ip}, and the dark photon \cite{Davoudiasl:2014kua}.

Recently, there have been proposed solutions to this anomaly through a scalar \cite{Chen:2015vqy} or a pseudoscalar Axion-Like Particle (ALP)\cite{Marciano:2016yhf} in a general framework. In this short paper, I will investigate the viability of these solutions, explore the relevant experimental limits, and highlight the experimental probes for their discovery for the mass range of their validity.

For the case of a pseudoscalar, the effective interaction with photons and fermions can be parametrized by:\footnote{This interaction can be viewed as an effective theory of a UV-complete model. One possible UV-completion that is consistent with the SM EW theory was introduced in \cite{Chen:2015vqy}. Such a model could lead to lepton-flavor violation through terms like $Y_{\phi ij} \phi \bar{l}_{i}l_{j}, i \neq j$. We ignore this possibility in this paper as it will not affect the interaction in Eq. \ref{eq:ScalarInteraction} and as it was studied in detail in \cite{Chen:2015vqy}. The interested reader is instructed to refer to \cite{Chen:2015vqy} for detailed analysis.}
\begin{equation}\label{eq:ScalarInteraction}
\mathcal{L} = \frac{1}{4} g_{\phi\gamma\gamma} \phi F_{\mu\nu}\tilde{F}^{\mu\nu} + i Y_{\phi ll} \phi \bar{\psi}  \gamma_{5} \psi,
\end{equation}
where $g_{\phi\gamma\gamma}$ is a dimensionful coupling, $Y_{\phi ll}$ is a dimensionless Yukawa coupling and $F^{\mu\nu}$, $\tilde{F}^{\mu\nu}$ are the magnetic field strength tensor and its dual respectively. For a scalar, $\tilde{F}^{\mu\nu}$ is replaced with $F^{\mu\nu}$ and there is no $i \gamma_{5}$ in the second term. Since $|\frac{\Delta a_{e}}{\Delta a_{\mu}}| \sim \frac{Y_{\phi ee}^{2}}{Y_{\phi \mu\mu}^{2}}$ (see Eqs. \ref{eq:PseudoLO} and \ref{eq:ScalarLO} below), and we can see from Eqs. \ref{eq:muDiscrepancy} and \ref{eq:eDiscrepancy} that within the allowed range of uncertainties, we could obtain $|\frac{\Delta a_{e}}{\Delta a_{\mu}}| \approx (\frac{m_{e}}{m_{\mu}})^{4}$ \footnote{Notice that from Eqs. \ref{eq:PseudoLO} and \ref{eq:ScalarLO}, $\Delta a_{l} \sim Y_{\phi ll}^{2} r^{-2}$, where $r=m_{\phi}/m_{l}$. The assumption in Eq. \ref{eq:YukawaCouplings} makes $\Delta a_{l} \sim m_{l}^{4}$, and thus $\frac{\Delta a_{e}}{\Delta a_{\mu}} \approx (\frac{m_{e}}{m_{\mu}})^{4}$.}, then we are motivated to define the Yukawa couplings to be proportional to the lepton mass:
\begin{equation}\label{eq:YukawaCouplings}
Y_{\phi ll} = \frac{m_{l}}{v} \equiv m_{l} g_{\phi ll},
\end{equation}
where $v \equiv g_{\phi ll}^{-1}$ is some model-dependent energy scale that is universal for all leptons, such as the axion decay constant or the radion constant. The coupling of the form given in Eq. \ref{eq:YukawaCouplings} has the additional advantage in that it arises in many UV-comepltions as Psuedo Nambu Goldstone Bosons (PNGB), such as axion models \cite{Hagmann}, Left-Right Twin Higgs Models \cite{Liu:2018pdq} and dark matter models with a scalar portal to the dark sector \cite{Nomura:2008ru} (also see \cite{Essig:2010gu}). PNGBs have an approximate shift symmetry and their interactions are proportional to some universal symmetry breaking scale. In addition, such models have been widely discussed in literature as an effective theory for solving the $g-2$ anomaly (see for example \cite{Chen:2015vqy}, \cite{Marciano:2016yhf}, \cite{Chen:2017awl}). I will focus on this form of Yukawa couplings throughout this paper.

It was shown in \cite{Chen:2015vqy} that the discrepancy in $g_{\mu}-2$ can be explained by a scalar with $Y_{\phi \mu\mu} \sim O(10^{-3})$, while in \cite{Marciano:2016yhf}, it was shown that an ALP pseudoscalar can explain both of the electron and the muon anomalies by considering the NLO contributions. 

\begin{figure}[!ht] 
  \centering
  \begin{minipage}[b]{0.18\textwidth}
    \includegraphics[width=\textwidth]{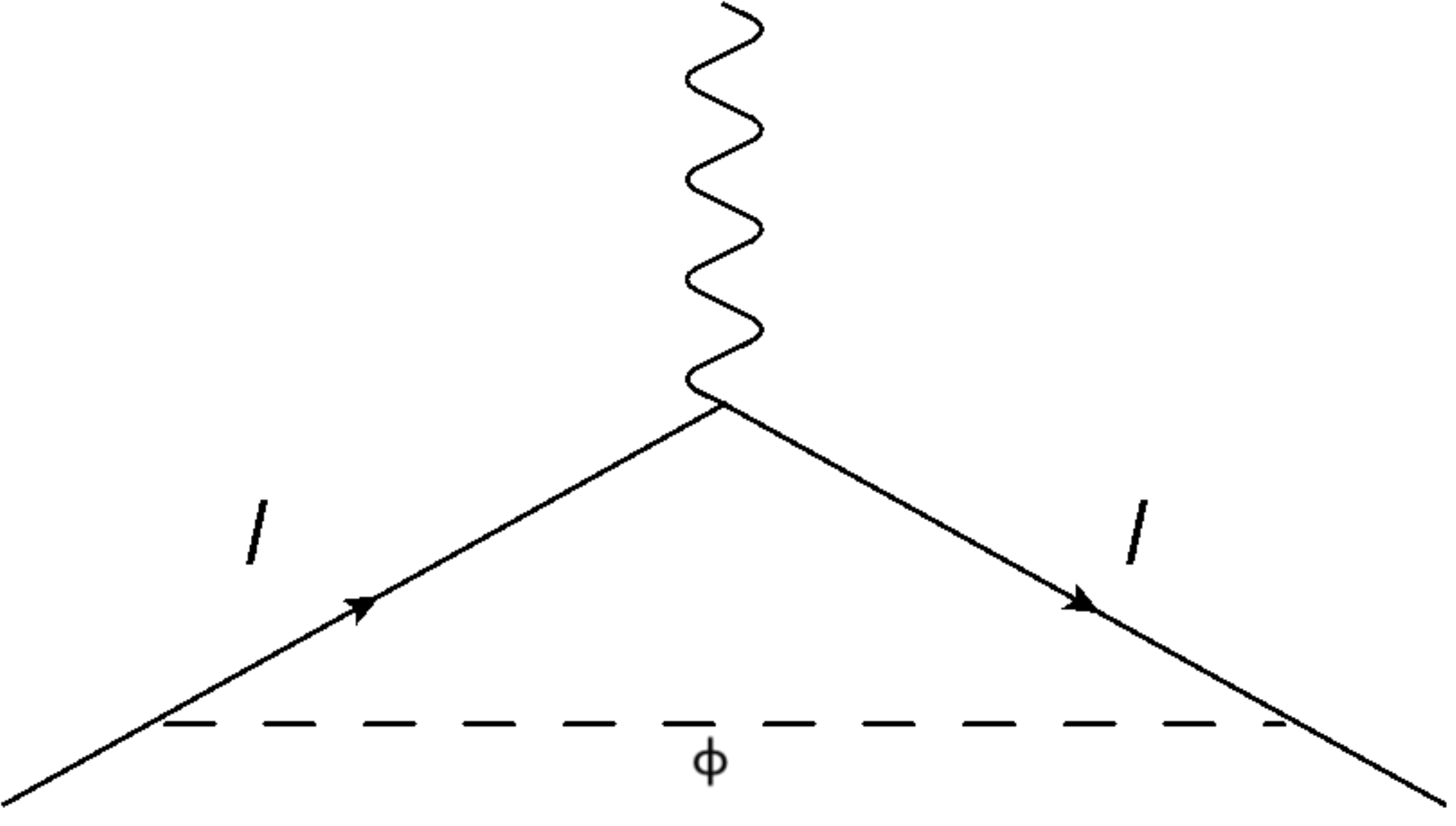}
  \end{minipage}
  \hspace{0.03cm}
  \begin{minipage}[b]{0.18\textwidth}
    \includegraphics[width=\textwidth]{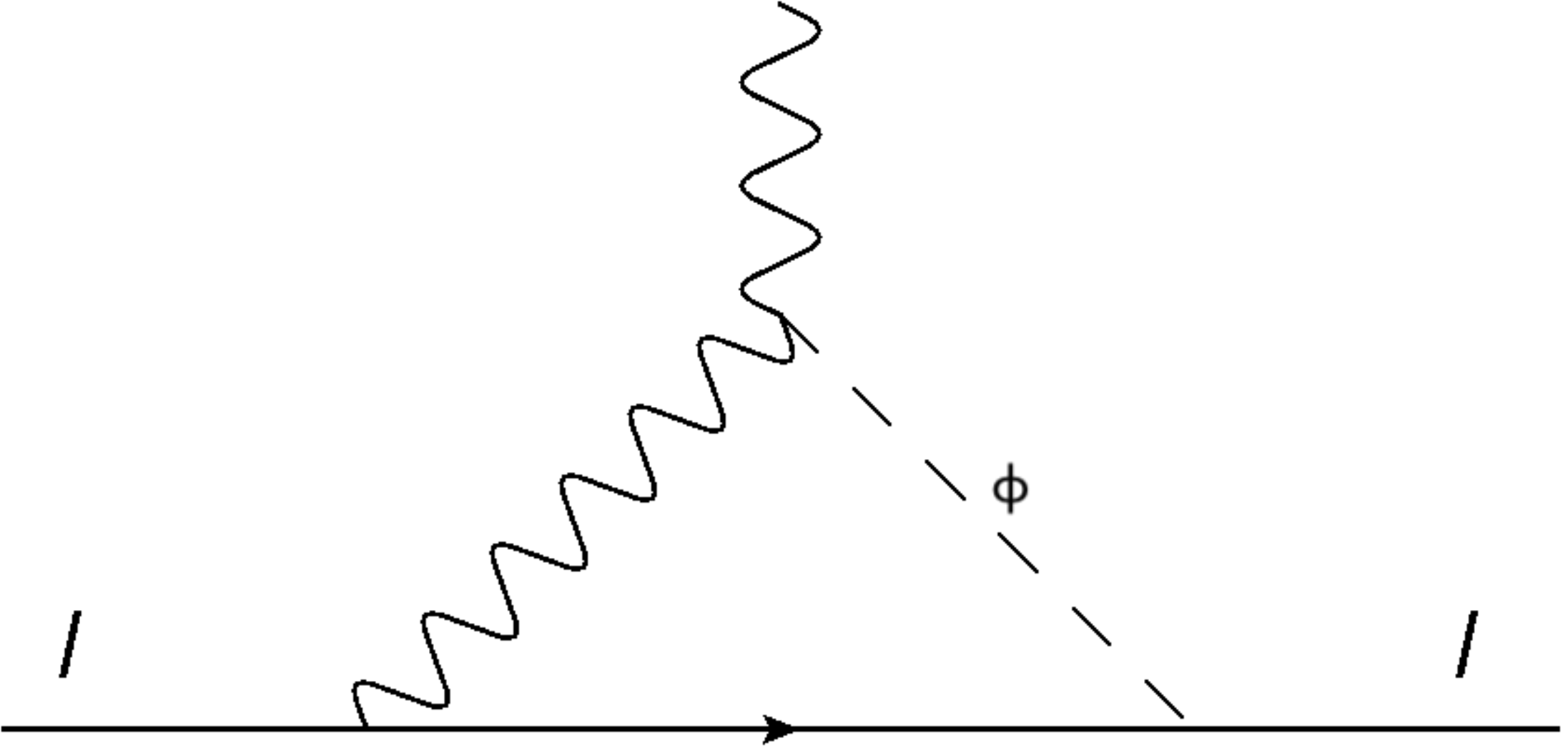}
  \end{minipage}
 \hspace{0.03cm}
\begin{minipage}[b]{0.18\textwidth}
    \includegraphics[width=\textwidth]{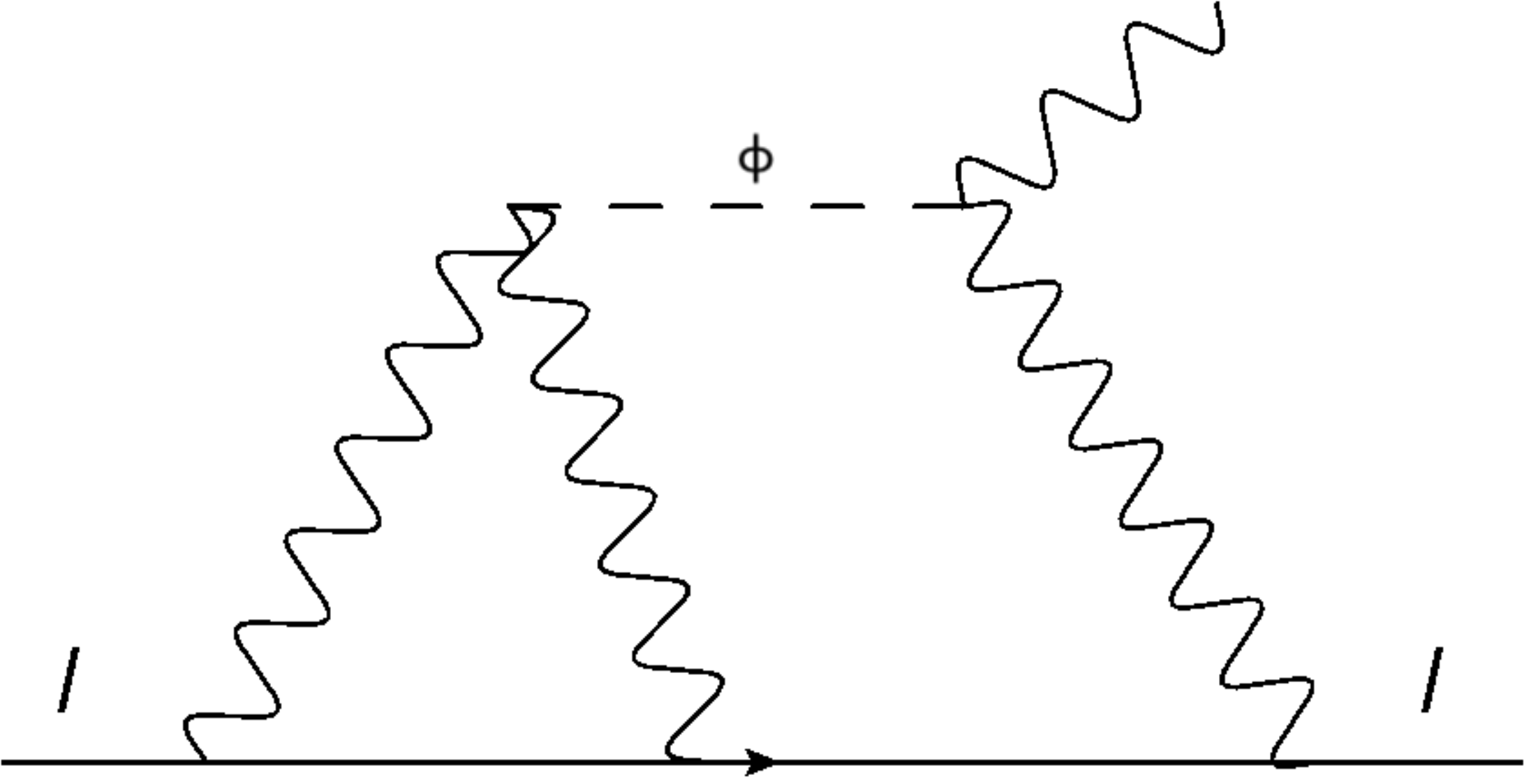}
  \end{minipage}  
\hspace{0.03cm}
  \begin{minipage}[b]{0.18\textwidth}
    \includegraphics[width=\textwidth]{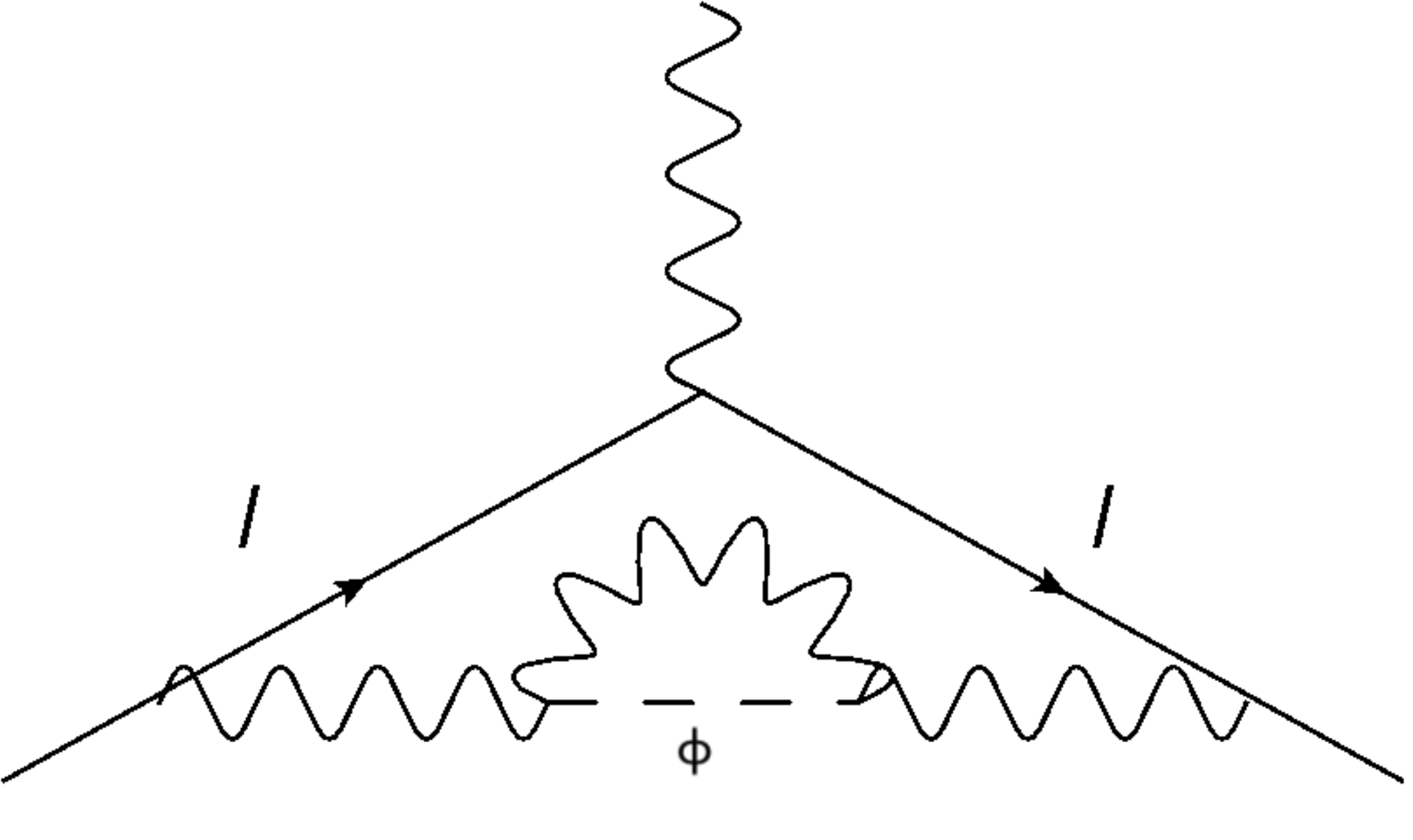}
  \end{minipage}
      \caption{LO (top left diagram) and NLO scalar/pseudoscalar contributions to the lepton anomalous magnetic moment.}
      \label{fig:g-2}
\end{figure}

The LO and NLO contributions to the $\Delta a_{\mu, e}$ are shown in Fig. (\ref{fig:g-2}). The LO contribution for the scalar, as well as for the pseudoscalar, was calculated in \cite{Chen:2015vqy}:
\begin{equation}\label{eq:PseudoLO}
\Delta a_{l}^{p} =-\frac{Y_{\phi ll}^{2}}{8\pi^{2}} r^{-2} \int_{0}^{1}dz \frac{(1-z)^{3}}{r^{-2}(1-z)^{2}+z},
\end{equation}
\begin{equation}\label{eq:ScalarLO}
\Delta a_{l}^{s} = \frac{Y_{\phi ll}^{2}}{8\pi^{2}} r^{-2} \int_{0}^{1} dz \frac{(1+z)(1-z)^{2}}{r^{-2}(1-z)^{2}+z},
\end{equation}
where $r \equiv \frac{m_{\phi}}{m_{l}}$. On the other hand, the NLO contribution includes the Barr-Zee (BZ) contribution (top right diagram in Fig. \ref{fig:g-2}), the two-loop Light-By-Light (LBL) contribution (bottom left diagram in Fig. \ref{fig:g-2}), and the Vacuum Polarization (VP) contribution (bottom right diagram in Fig. \ref{fig:g-2}). These contributions are the same for both the scalar and the pseudoscalar cases and are given by\cite{Marciano:2016yhf}:
\begin{equation}\label{eq:BZ}
a_{l,\phi}^{\text{BZ}} \simeq \Big(\frac{m_{l}}{4\pi^{2}} \Big) g_{\phi \gamma\gamma} Y_{\phi ll} \ln \frac{\Lambda}{m_{\phi}},
\end{equation}

\begin{equation}\label{eq:LBL}
a_{l,\phi}^{\text{LBL}} \simeq \frac{3\alpha}{\pi} \Big( \frac{m_{l}g_{\phi \gamma \gamma}}{4\pi} \Big)^{2} \ln^{2} \frac{\Lambda}{m_{\phi}},
\end{equation}

\begin{equation}\label{eq:VP}
a_{l,\phi}^{\text{VP}} \simeq \frac{\alpha}{\pi} \Big(  \frac{m_{l} g_{\phi \gamma\gamma}}{12\pi} \Big)^{2} \ln \frac{\Lambda}{m_{\phi}},
\end{equation}
where $\phi$ is either a scalar or a pseudoscalar, $g_{\phi \gamma\gamma}$ is the dimensionful coupling of  $\phi$ to photons, and $\Lambda$ is some UV cutoff scale that is assumed to be much larger than $m_{\phi}$. I will set the cutoff scale $\Lambda =1$ TeV throughout this paper. Notice that since the Lagrangian in Eq. \ref{eq:ScalarInteraction} is CP-conserving, there will be no contribution to the lepton's Electric Dipole Moment (EDM). I will ignore the more general case where CP-violating terms are present.

\section{\label{sec:Favored}Favored Region}
In this section, I will investigate the parameter space and try to establish the favored region for both the scalar and the pseudoscalar cases. If we inspect equations \ref{eq:PseudoLO} through \ref{eq:VP}, we notice the following:

\begin{itemize}
\item For a scalar, the LO contribution is always positive, while for a pseudoscalar it is negative. As for the NLO contributions, we can see that the LBL and the VP are always positive, while the BZ contribution depends on the sign of $g_{\phi\gamma\gamma}Y_{\phi ll}$.
\item Since the central measured anomaly for the muon is positive, while for the electron it is negative, a scalar solution can easily accommodate the muon anomaly. However, yielding the central measured electron anomaly would require the assumption that $g_{\phi\gamma\gamma}Y_{\phi ll} < 0$ so that the BZ contribution can offset all other (positive) contributions. This would require large (nonperturbative) Yukawa couplings for the electron and therefore it is disfavored. Thus I will assume that all couplings are positive for the scalar case. This means that the central measured electron anomaly cannot be produced. However, it is possible to show that the contribution is within $\Delta a_{e} +2\sigma_{e}$ \footnote{See added note at the end of this paper.}.
\item As the LO contribution for the pseudoscalar case is negative, it can easily accommodate the measured electron anomaly, however, in order to yield the (positive) measured muon anomaly, one needs somewhat large couplings to photons while keeping the Yukawa coupling somewhat small in order for the NLO contributions to dominate over the LO. Nonetheless, it is possible to find such solutions while maintaining perturbative couplings as we shall see below.
\item For the pseudoscalar case, if we assume that $g_{\phi\gamma\gamma}>0$, then $Y_{\phi ee} <0$ while $Y_{\phi\mu\mu}$ could be either positive or negative. On the other hand, assuming that $g_{\phi\gamma\gamma}<0$ yields exactly the same solution but with opposite signs.

\end{itemize}

\subsection{\label{sec:PSFavored}Pseudoscalar}
\begin{figure}[!h] 
  \centering
    \includegraphics[width=0.4\textwidth]{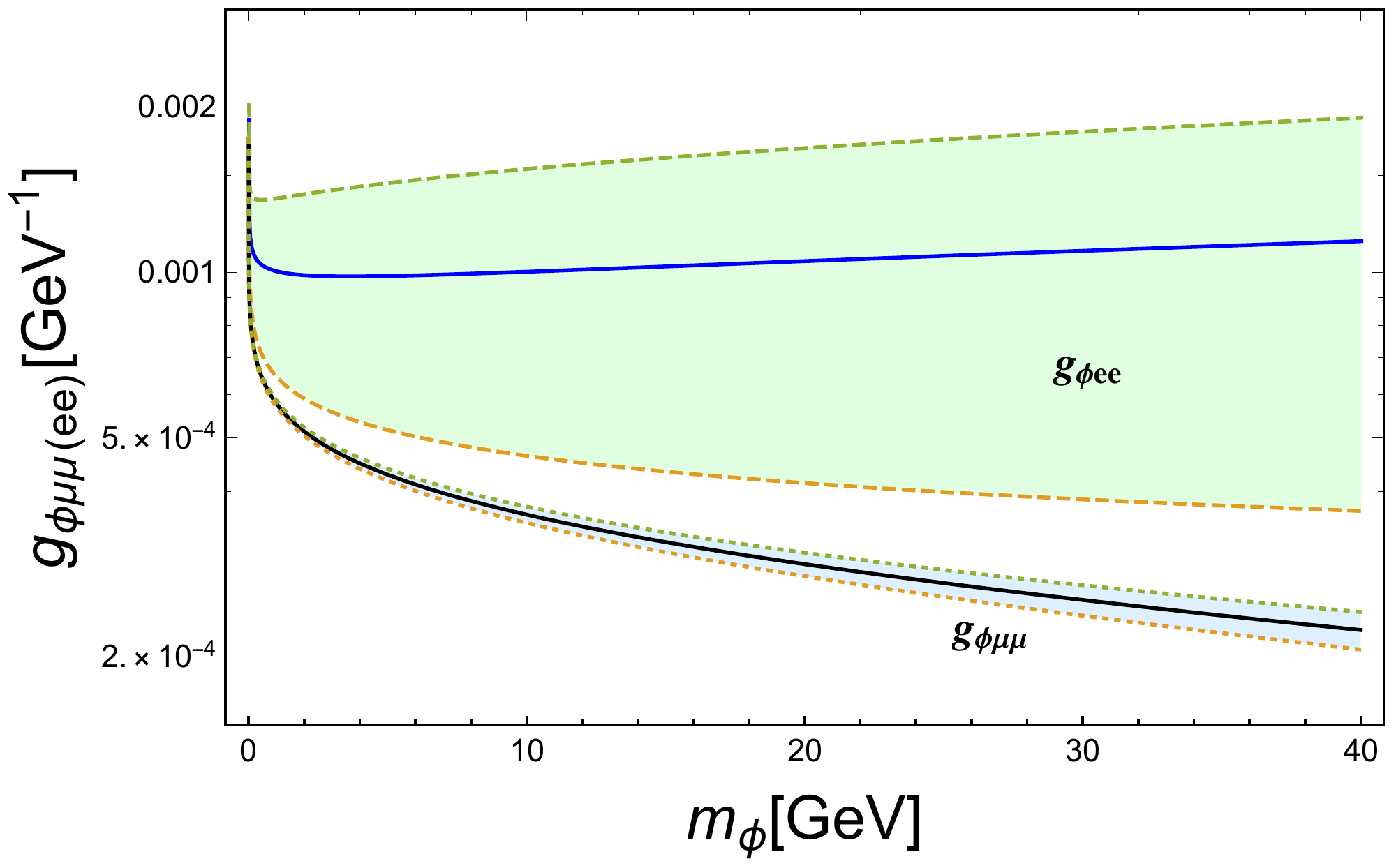}
      \caption{The $2\sigma$ allowed region for $e$ (green) and $\mu$ (blue) of the parameter space for the pseudoscalar solution assuming $g_{\phi \gamma\gamma} = -0.05$ GeV$^{-1}$. The lack of any overlap between the favored regions for the electron and the muon indicates that we cannot have a solution where Eq. \ref{eq:YukawaCouplings} is valid. Therefore the pseudoscalar solution is disfavored.}
      \label{fig:PseudoFavored}
\end{figure}

For concreteness, I will assume that $g_{\phi\gamma\gamma}<0$. Fig. (\ref{fig:PseudoFavored}) shows the $2\sigma$ allowed regions for $e$ and $\mu$ with $g_{\phi\gamma\gamma} = -0.05$ GeV$^{-1}$. Notice that there is no overlap between the two regions even for small masses, which means that there is no region in the parameter space where Eq. \ref{eq:YukawaCouplings} is valid. As a matter of fact, there is no pseudoscalar solution in the whole parameter space where Eq. \ref{eq:YukawaCouplings} is true. In addition, for both Yukawa couplings to have the same sign, one needs $|g_{\phi\gamma\gamma}| \gtrsim 0.03 \hspace{1mm} \text{GeV}^{-1}$, which is excluded by cosmological observations \cite{Jaeckel:2015jla}, therefore the pseudoscalar solution is disfavored. One can avoid these constraints by assuming that $Y_{\phi \mu \mu} >0$ and $Y_{\phi ee} <0$, however, one needs to justify this assumption. I will disregard the pseudoscalar solution in the remainder of this paper.

\subsection{\label{sec:SFavored}Scalar}
With the assumption that all couplings are positive, I attempt at finding the favored region in the parameter space for the scalar case. Here, I will only focus on solutions of the form in Eq. \ref{eq:YukawaCouplings} and discard all other possibilities. 

Notice that we have three parameters, namely $m_{\phi}$, $g_{\phi ll}$ and $g_{\phi \gamma\gamma}$. I will fix $g_{\phi\gamma\gamma}$ and keep $g_{\phi ll}$ and $m_{\phi}$ as free parameters. I will select appropriate benchmark points for $g_{\phi\gamma\gamma}$ by minimizing the $\chi^{2}$ of the electron and muon measurements:
\begin{equation}\label{eq:ChiSq}
\chi^{2} = \frac{(\Delta_{\mu}^{\text{Exp}}-\Delta_{\mu}^{\phi})^{2}}{\sigma_{\mu}^{2}} +  \frac{(\Delta_{e}^{\text{Exp}}-\Delta_{e}^{\phi})^{2}}{\sigma_{e}^{2}}.
\end{equation}

More concretely, I fix the value of $g_{\phi \gamma\gamma}$, then I use Eq. \ref{eq:ChiSq} to find $g_{\phi ll}$ and $m_{\phi}$ that minimize $\chi^{2}$, together with the value of $\chi^{2}$ at the minimum.  Then I will scan through a wide range of $g_{\phi \gamma\gamma}$, and then set the benchmark points where $\chi^{2}_{min}$ is the smallest.

\begin{figure}[!ht] 
  \centering
    \includegraphics[width=0.4\textwidth]{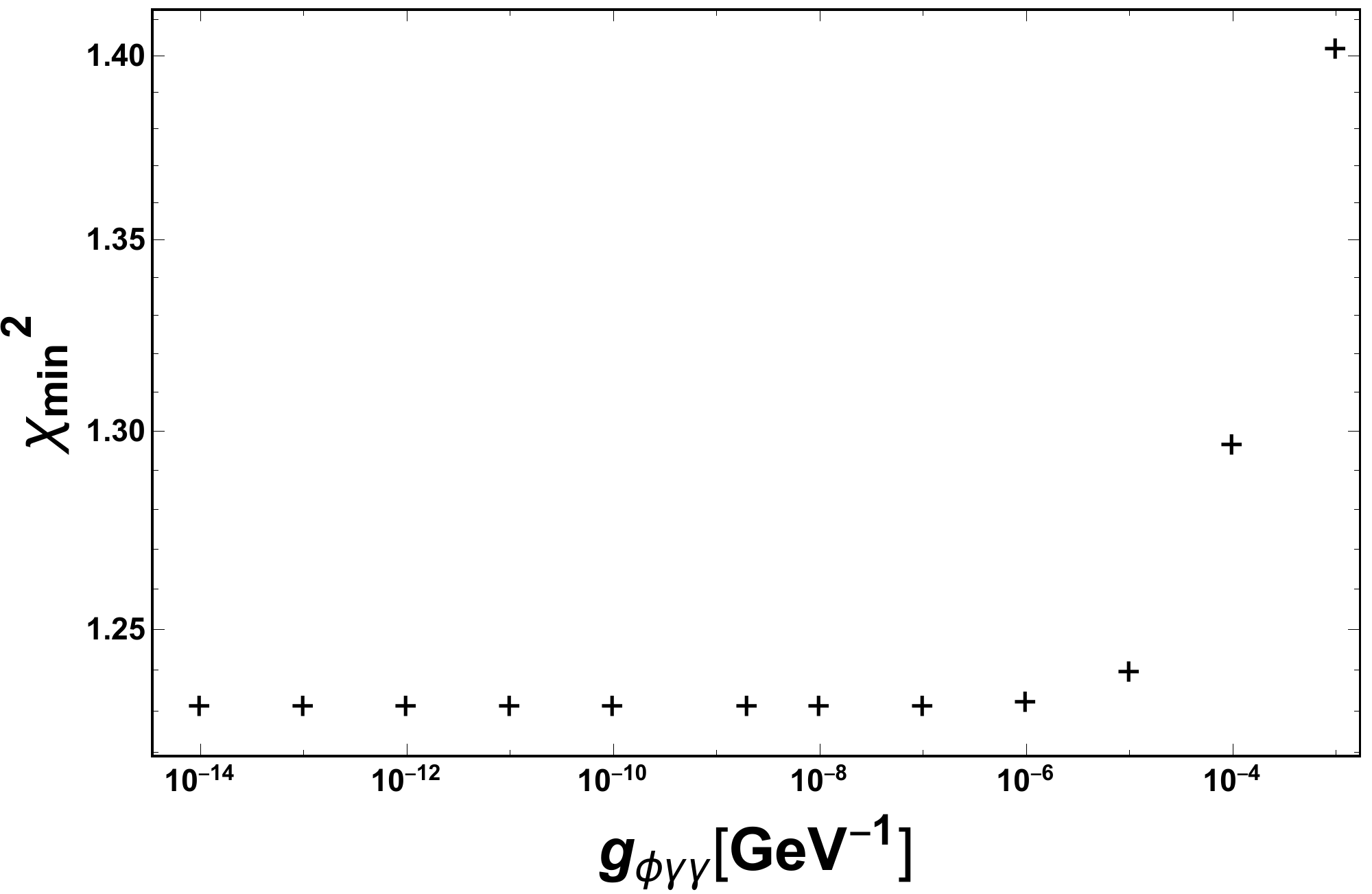}
      \caption{$\chi^{2}_{min}$ vs. $g_{\phi\gamma\gamma}$. The plot shows that smaller couplings to photons are favored.}
      \label{fig:ChiVg}
\end{figure}

Fig. \ref{fig:ChiVg} shows $\chi^{2}_{min}$ for several values of $g_{\phi \gamma\gamma}$. As the plot shows, $g_{\phi \gamma\gamma} \lesssim 10^{-6}$ GeV$^{-1}$ yields the lowest values of $\chi^{2}$. Notice that $\chi^{2}$ becomes almost constant for smaller couplings. This is reasonable as when the coupling to photons becomes very small, the NLO contributions become negligible and the LO contribution is dominant. This high-level analysis seems to favor smaller couplings to photons, suggesting that the coupling to leptons is the dominant coupling. This is consistent with the cosmological constraints on ALPs (see for instance \cite{Jaeckel:2015jla}). I will focus on this scenario and I will choose $g_{\phi \gamma\gamma} = 10^{-6}$ and $10^{-11}$ GeV$^{-1}$ as two benchmark points. Notice that to a good level of accuracy, the second benchmark point is representative of the entire region of $g_{\phi \gamma\gamma} \lesssim 10^{-8}$ GeV$^{-1}$, with very similar favored regions for the predicted mass and coupling.

\begin{figure}[!ht] 
  \centering
  \begin{minipage}[b]{0.4\textwidth}
    \includegraphics[width=\textwidth]{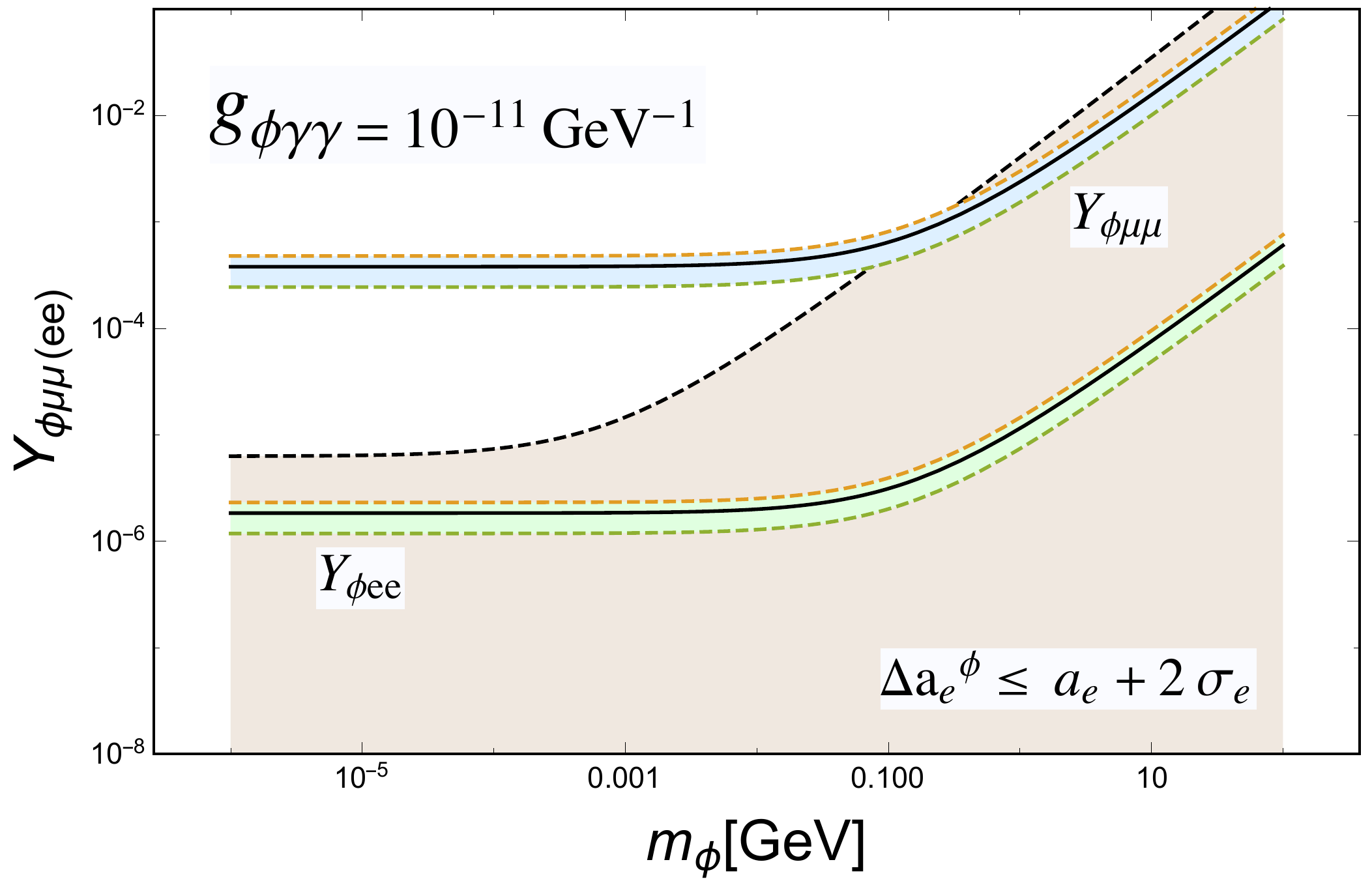}
  \end{minipage}
  \hspace{0.01cm}
  \begin{minipage}[b]{0.4\textwidth}
    \includegraphics[width=\textwidth]{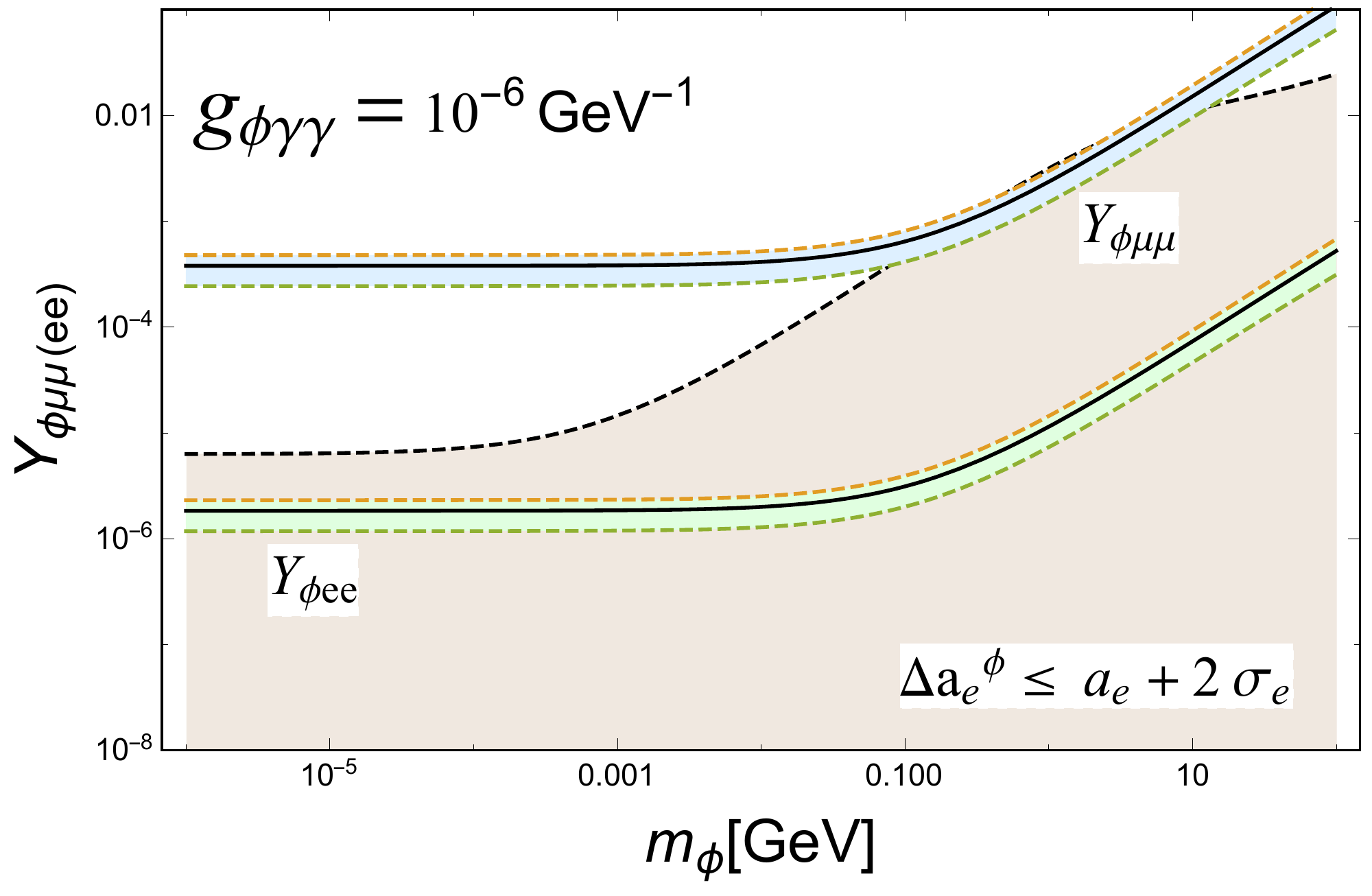}
  \end{minipage}
      \caption{The $2\sigma$ bands corresponding to the allowed region for $e$ and $\mu$ for the scalar case at the two benchmark points with $g_{\phi\gamma\gamma} = 10^{-11}$ GeV$^{-1}$ (top) and $10^{-6}$ GeV$^{-1}$ (bottom). The brown region corresponds to $\Delta a_{e}^{\phi} \leq a_{e}+2\sigma_{e}$. Notice here that for the blue and green bands $\frac{Y_{\phi \mu\mu}}{Y_{\phi ee}} = \frac{m_{\mu}}{m_{e}}$ in accordance with Eq. \ref{eq:YukawaCouplings}.}
      \label{fig:AllowedRegion}
\end{figure}

Now we can use Eqs. \ref{eq:ScalarLO} - \ref{eq:VP} in order to find the allowed region in the $m_{\phi} - Y_{\phi ll}$ parameter space corresponding to a $2\sigma$ deviation from the central values. In order to set an upper limit on $m_{\phi}$, we demand that all Yukawa couplings remain perturbative. Obviously, the most stringent bound comes from $Y_{\phi\tau\tau}$ as it has the largest value. Requiring that $Y_{\phi\tau\tau} \lesssim 1$, we obtain an upper bound on $m_{\phi}$ of $\sim 45 (50)$ GeV for $g_\phi\gamma\gamma = 10^{-11}(10^{-6}$ GeV). Fig. (\ref{fig:AllowedRegion}) shows the allowed region corresponding to the two benchmark points. The plots show the $2\sigma$ bands for the Yukawa couplings to electrons and muons assuming that Eq. \ref{eq:YukawaCouplings} holds. In addition, the plots also show the region where the contribution to $\Delta a_{e}$ is within $2\sigma$ of the measured value assuming that the Yukawa couplings to leptons are independent of one another. Notice that the brown region corresponds to $\Delta a_{e}^{\phi} \in (0,\Delta a_{e}^{\text{Exp}}+2\sigma_{e}]$ since we are assuming positive couplings. Thus as noted earlier, there is no point in the parameter space that can yield the central value of the measured $\Delta a_{e} =-91 \times 10^{-14}$.

\begin{figure}[!ht] 
  \centering
  \begin{minipage}[t][5cm][t]{0.4\textwidth}
    \includegraphics[width=\textwidth, height=0.85\linewidth]{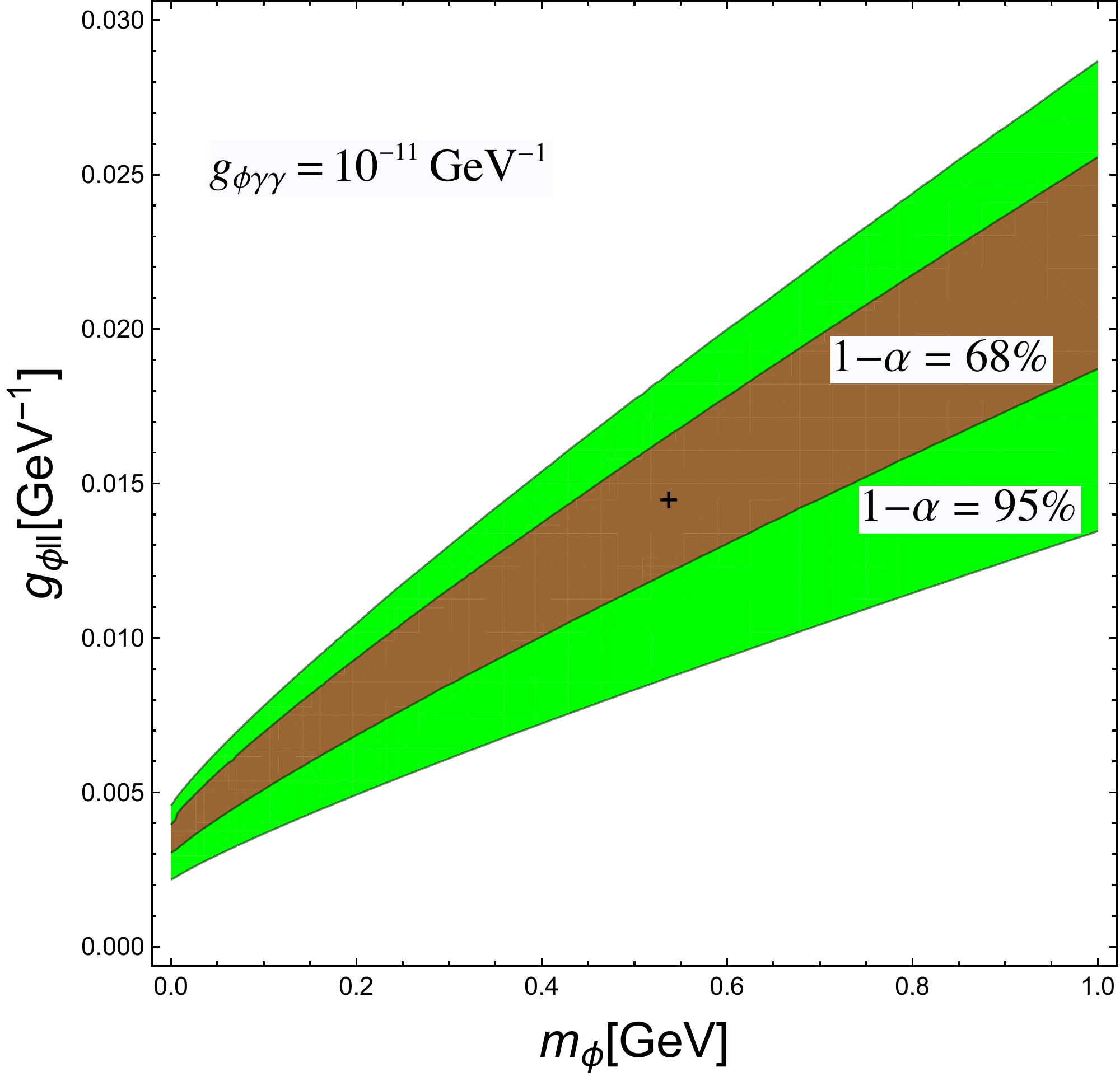}
  \end{minipage}
  \begin{minipage}[t][5cm][t]{0.4\textwidth}
    \includegraphics[width=\textwidth, height=0.85\linewidth]{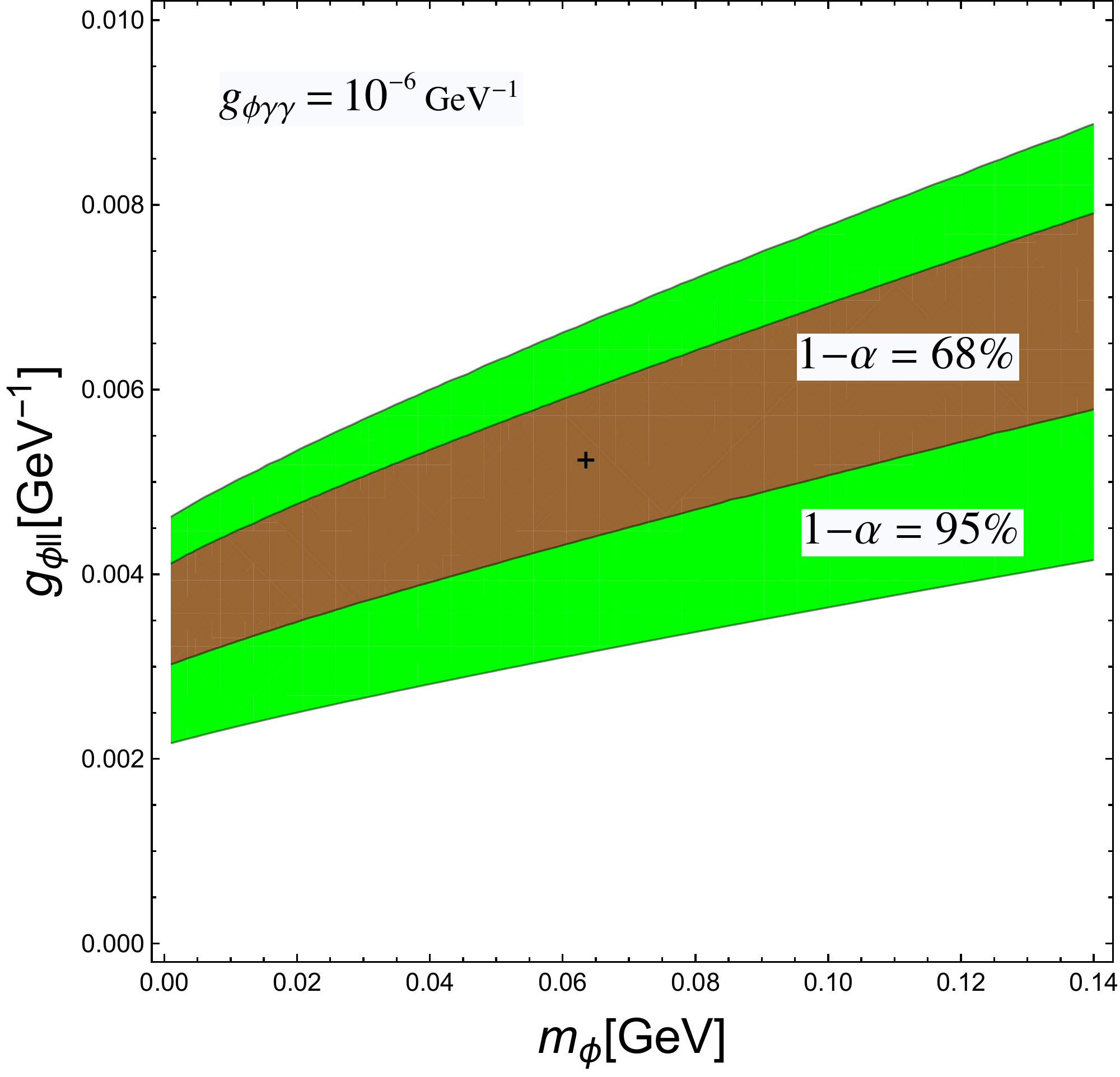}
  \end{minipage}
      \caption{The favored point in the $m_{\phi} - g_{\phi ll}$ parameter space of the scalar solution, with $68\%$ and $95\%$ confidence level contours for $g_{\phi\gamma\gamma} = 10^{-11}$ GeV$^{-1}$ (top) and $10^{-6}$ GeV$^{-1}$ (bottom).}
      \label{fig:FavoredRegion}
\end{figure}

We can carry the $\chi^{2}$ analysis further to find the favored region in the  $m_{\phi} -g_{\phi ll}$ parameters space for the benchmark points. Fig. (\ref{fig:FavoredRegion}) shows the scalar mass and coupling to leptons that minimize $\chi^{2}$, together with the $68\%$ and $95\%$ confidence level contours. Notice that this region is a subset of the allowed region in Fig. (\ref{fig:AllowedRegion}). For $g_{\phi\gamma\gamma} = 10^{-11}$ GeV$^{-1}$, we find a predicted scalar mass of $\sim 540$ MeV, with $g_{\phi ll} \simeq 1.45 \times 10^{-2}$ GeV$^{-1}$. For this point, one finds a predicted anomaly $\Delta a_{\mu(e)} = 273 \times 10^{-11} (8 \times 10^{-18})$. On the other hand, for the second benchmark point ($g_{\phi \gamma\gamma} = 10^{-6}$ GeV$^{-1}$), the predicted mass and coupling are $64$ MeV and $5 \times 10^{-3}$ GeV$^{-1}$ respectively, with $\Delta a_{\mu(e)} = 272 \times 10^{-11} (5 \times 10^{-17})$. 
This is consistent with the results found in \cite{Chen:2015vqy}.

Notice that the predicted values of the electron anomaly for these benchmark points are much smaller that the (absolute) measured central value of $91 \times 10^{-14}$ (see Eq. \ref{eq:eDiscrepancy}), albeit they are still within the $2\sigma$ limit. This is logical as we can see from Eq. \ref{eq:ScalarLO}, $\Delta a_{e} \sim \Delta a_{\mu} \times \big( \frac{m_{e}}{m_{\mu}} \big)^{4}$ if the Yukawa coupling has the form given in Eq. \ref{eq:YukawaCouplings}. This is an important prediction to test this model. That is, we claim that if indeed this model is correct, then more accurate measurements of electron anomaly should yield $ \Delta a_{e} \sim \Delta a_{\mu} \times \big( \frac{m_{e}}{m_{\mu}} \big)^{4} \sim O(10^{-18})$. If future measurements of the electron anomaly are inconsistent with this, then the assumption in Eq. \ref{eq:YukawaCouplings} would be ruled out and other explanations would be needed. Another important prediction of this model is the tau anomaly. Given that Eq. \ref{eq:YukawaCouplings} predicts the anomlay to be proportional to the mass, then  $(g_{\tau}-2) $ should be large enough to be measured. Specifically, we predict $(g_{\tau}-2) = 8(1.5) \times 10^{-6}$ for $g_{\phi \gamma\gamma} = 10^{-11}(10^{-6})$ GeV$^{-1}$.

\section{\label{sec:Probes}Experimental Probes and Limits}

For the mass range $\lesssim 50$ GeV, the most relevant constraints come from the BaBar experiment, beam dump experiments, the $Z$ decay, the LEP mono-$\gamma$ searches, $ee \rightarrow \tau\tau$ and $ee \rightarrow \tau\tau+\gamma$ searches, and from the bounds on solar emission and the on emission from HB stars. Although there is overlap between the results of this chapter and \cite{Batell:2016ove}, where they discuss a similar effective model and a UV completion through the 2HDM, there are several novel features in this work, including bounds from the Z decay, LEP searches and the solar and HB stars. In addition, here we  attempt at explaining both the electron and the muon anomalies and we extend the range of the mass over the entire range of validity. Thus our work should be viewed as complementary to theirs.

\subsection{\label{sec:BaBar}BaBar}
\begin{figure}[!ht] 
  \centering
    \includegraphics[width=0.4\textwidth]{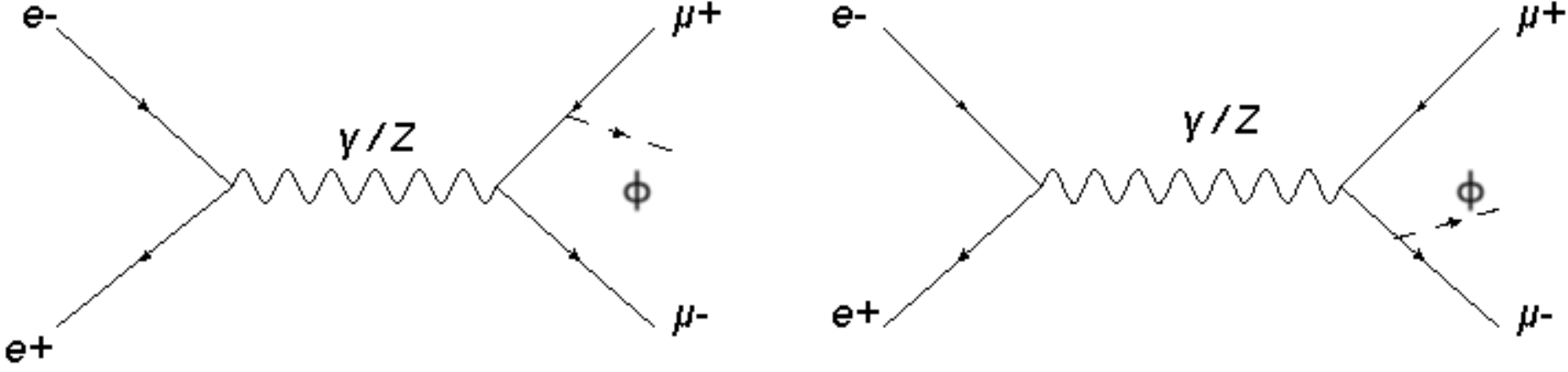}
      \caption{The dominant tree-level contributions to $e^{+}e^{-} \rightarrow \mu^{+}\mu^{-} \phi$.}
      \label{fig:eemumuSigma}
\end{figure}

Recent results from the BaBar experiment \cite{TheBABAR:2016rlg} searching for the process $e^{+}e^{-} \rightarrow \mu^{+}\mu^{-}Z', Z' \rightarrow \mu^{+}\mu^{-}$ can be important for constraining the parameters space. The results can be used to extract the constraints on the process $e^{+}e^{-} \rightarrow \mu^{+}\mu^{-}\phi, \phi \rightarrow \mu^{+}\mu^{-}$ \footnote{\cite{Bauer:2017ris} extracts the constraints from BaBar's results for the pseudoscalar case.}. The tree-level Feynman diagrams that contribute to this process are shown in Fig. (\ref{fig:eemumuSigma}), where I have neglected the diagrams where the scalar is radiated by the initial state particles or by the intermediate particle, since the coupling to muons dominates over the coupling to electrons and photons, and I am assuming a subleading coupling to the $Z$. 
Fig. (4) in \cite{TheBABAR:2016rlg} sets an upper limit $\sigma_{\text{Max}}$ on the process $e^{+}e^{-} \rightarrow \mu^{+}\mu^{-}Z', Z' \rightarrow \mu^{+}\mu^{-}$. Therefore, we can extract the limits on the $g_{\phi ll}-m_{\phi}$ parameter space by requiring:
\begin{equation}\label{eq:BaBarLimit}
\sigma(e^{+}e^{-} \rightarrow \mu^{+}\mu^{-}\phi) \times \text{Br} (\phi \rightarrow \mu^{+}\mu^{-}) < \sigma_{\text{Max}}. 
\end{equation}

The excluded part of the parameter space is shown in the light gray region in Fig. (\ref{fig:limits}). Notice that it does not constrain the favored region corresponding to a $2\sigma$ deviation, although larger deviations would be constrained. This result also shows us that any new searches from BaBar (or similar experiments like Belle II) need to be at least an order of magnitude better in order to explore the favored region in the parameter space, or that searches should be made for a similar process with taus instead of muons, as the former has a larger coupling. Thus, there is a good motivation to search for processes like $ee \rightarrow 4\tau$.

Notice that in calculating the branching fraction, we are assuming that the scalar can only decay to leptons or photons. If the scalar is allowed decay to other SM particles, the limits will become even weaker. Also notice that the upper limit in BaBar's results found in Fig. (4) of \cite{TheBABAR:2016rlg} is spiky, therefore the upper limit used in Eq. \ref{eq:BaBarLimit} contains significant uncertainty of about a factor of a few. In extracting $\sigma_{\text{Max}}$, I was conservative and used the smallest cross-section. A less conservative estimate would relax the limits further.

\subsection{\label{sec:Zdecay}$Z$ Decay\footnote{\cite{Lebedev:1999vc,Lebedev:2000ix} also discuss the 1-loop corrections to the $Z$-decay in the context of R-parity violating extensions to SUSY and in the context of type-II 2HDM.}}
The excellent measurements of the $Z$ decay width and branching fractions present us with a potentially suitable tool for probing the parameter space. In particular, we can explore the limits associated with the scalar loop correction to the $Z$ decay to a pair of leptons.

\begin{figure}[!ht] 
  \centering
    \includegraphics[width=0.4\textwidth]{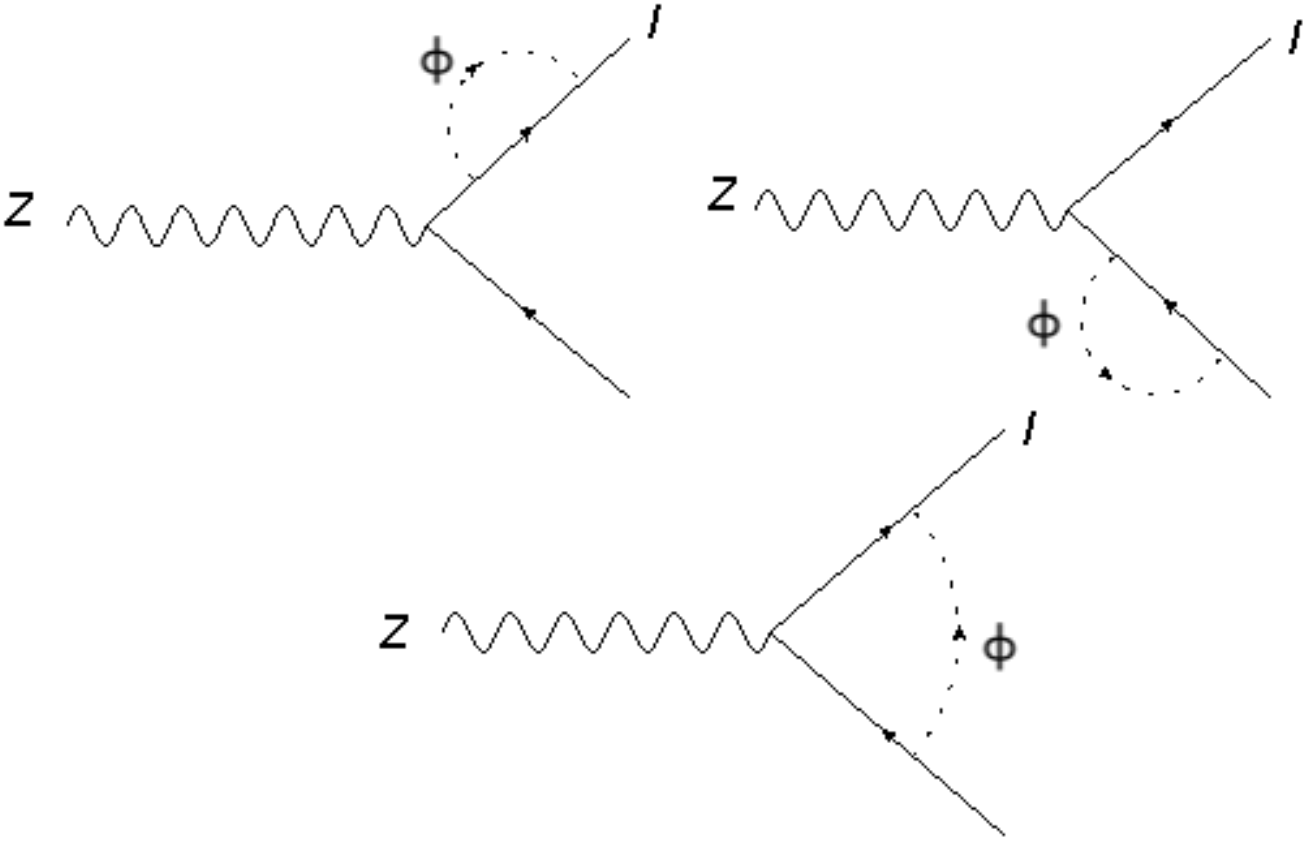}
      \caption{Scalar NLO corrections to $Z \rightarrow l\bar{l}$.}
      \label{fig:Loop}
\end{figure}

The scalar loop corrections can significantly affect the leptonic decay width of the $Z$ boson. The NLO corrections to $Z \rightarrow l\bar{l}$ are shown in Fig. (\ref{fig:Loop}), where the coupling of $\phi$ to $Z$ is assumed to be subdominant compared with the coupling to leptons. Notice here that UV divergences in the leg corrections cancel that in the vertex correction, and that for a massive $\phi$ the result is free of IR divergences.

Dropping the lepton mass in the loops, and keeping $m_{\phi}$ only as an IR regulator, which is justified for $m_{\phi} \gg m_{l}$ where this bound is moslty relevant, the NLO correction is approximately given by:

\begin{equation}\label{eq:loop}
\delta \Gamma (Z \rightarrow l\bar{l}) \simeq - \Gamma_{0} \frac{m_{l}^{2} g_{\phi ll}^{2}}{8\pi^{2}} \Bigg[ \log{\Big( \frac{M_{Z}^{2}}{m_{\phi}^{2}} \Big)}  -2\Bigg],
\end{equation}
where $\Gamma_{0}$ is the LO decay width given by:

\begin{equation}\label{eq:tree}
\Gamma_{0} (Z \rightarrow l\bar{l}) = \frac{g^{2}(g_{V}^{2}+g_{A}^{2})}{48\pi \cos^{2}{\theta_{W}}} M_{Z} \sqrt{1-\frac{4m_{l}^{2}}{M_{Z}^{2}}}.
\end{equation}

We can compare this correction with the branching fractions of the leptonic $Z$ decays, which are given by \cite{Patrignani:2016xqp}:

\begin{equation}\label{eq:Ztoee}
\text{Br}(Z \rightarrow e^{+}e^{-}) = (3363.2 \pm 4.2) \times 10^{-3} \hspace{1mm}\%,
\end{equation}
\begin{equation}\label{eq:Ztomumu}
\text{Br}(Z \rightarrow \mu^{+}\mu^{-}) = (3366.2 \pm 6.6) \times 10^{-3} \hspace{1mm}\%,
\end{equation}
\begin{equation}\label{eq:Ztotautau}
\text{Br}(Z \rightarrow \tau^{+}\tau^{-}) = (3369.6 \pm 8.3) \times 10^{-3} \hspace{1mm}\%.
\end{equation}

As it turns out, the decay to $\tau\tau$ provides the most stringent constraints in spite of the larger uncertainty in its branching fraction. This is due to its larger coupling given the assumption in Eq. \ref{eq:YukawaCouplings}.

The excluded region of the parameter space at a $2\sigma$ level is shown in red in Fig. (\ref{fig:limits}). As can be seen from the plots, there is some tension between the $Z$-loop decay and the allowed region near $m_{\phi} \sim 10$ GeV, although the alowed region is not fully excluded. Notice here that since we are dropping the mass of the tau, the bound will be less reliable for the range $m_{\phi} < m_{\tau}$, however, this is unimportant as in that region the $Z$-decay constraint is far from the favored region and therefore our results are unaffected. On the other hand, in the region of interest where $m_{\phi}$ is larger than a few GeV, dropping the mass of the tau is justified and won't impact the results significantly. The same argument applies for the constraint obtain from $e^{+}e^{-} \rightarrow \tau^{+}\tau^{-}$ we discuss below.

\subsection{\label{sec:eetautau}Constraints from $e^{+}e^{-} \rightarrow \tau^{+}\tau^{-}$}
The analysis conducted in the previous section can be extended to the $e^{+} e^{-} \rightarrow \tau^{+}\tau^{-}$ searches conducted by the LEP experiment \cite{Abbiendi:2003dh, Abbiendi:2000hu, Abbiendi:1999wm, Acciarri:1994gx, Akrawy:1990tn, Acton:1993yc, Akrawy:1989gf, Abreu:2000mh, Akrawy:1990dg, Aarnio:1990qe, Abreu:1991wj, Buskulic:1993gu, Decamp:1990ky, Adriani:1993ca, Decamp:1991aj, Buskulic:1996ua}, the KEK collaboration \cite{Adachi:1987gr, Band:1989av, Velissaris:1994rv, Howell:1992ar}, DESY-PETRA collaboration \cite{Behrend:1982gg, Braunschweig:1989bw, Bartel:1985mh, Hegner:1989rd, Behrend:1989wc, Bartel:1985cs}, and the SLAC-PEP experiment \cite{Gan:1985st}. 

Here, the process of $ee \rightarrow \tau\tau$ proceeds through the s-channel with a photon or a $Z$ propagator, and the NLO loop correction due to $\phi$ will be identical to the case of the $Z$ decay shown in Fig. (\ref{fig:Loop}). So we can set the $2\sigma$ bound as:

\begin{equation}\label{eq:eetautauBound}
\delta \sigma_{\text{loop}}(ee \rightarrow \tau\tau) < 2\delta_{\sigma}^{\text{Measured}},
\end{equation}
where
\begin{equation}\label{eq:eetautauXsection}
\sigma_{\text{loop}} = \sigma_{0} \Bigg[1+ \frac{m_{\tau}^{2} g_{\phi ll}^{2}}{8\pi^{2}} \Big(2- \log{\Big(\frac{s}{m_{\phi}^{2}}\Big)}   \Big)\Bigg],
\end{equation}
and $\sigma_{0}$ is the tree level cross-section. The bound is shown in Fig. (\ref{fig:limits}) in light green. As the plot shows, and similar to the case of the $Z$ decay, there is significant tension with experimental results for $m_{\phi} \gtrsim 10$ GeV, although this region is not fully excluded.

\subsection{\label{sec:beamdump}Muon Beam Dump Experiments}
Muon beam dump experiments provide a powerful tool for probing the mass range $\sim 1 - 200$ MeV. The relevant constraints come from Orsay \cite{Davier:1989wz} and the E137 experiment at SLAC \cite{Bjorken:1988as} (also see \cite{Chen:2017awl} for a summary). 

In the Orsay beam dump experiment, searches for the light Higgs boson in the 2HDM were conducted through looking for the process $e N \rightarrow e N H, H \rightarrow e^{+}e^{-}$. In \cite{Davier:1989wz}, the coupling of the lighter Higgs to electrons is assumed to be:
\begin{equation}\label{eq:2HDMCoupling}
g_{hee} = \frac{m_{e}}{v}\tan\beta,	
\end{equation}
where $\tan \beta = v_{1}/v_{2}$, the ratio of the two doublets' VEVs. The results show the excluded region in the $\tan \beta - m_{h}$ parameter space. Therefore, they can readily be extrapolated to this model by setting $g_{\phi ll} \equiv \frac{\tan\beta}{v}$.

On the other hand, \cite{Bjorken:1988as} presents the results of the E137 beam dump searches for axions produced via bremsstrahlung followed by the subsequent decay to $e^{+}e^{-}$. The results were extracted for the case of a scalar in \cite{Chen:2017awl}, so I will just use their results.

I show these constraints in magenta (Orsay) and orange (E137) in Fig. (\ref{fig:limits}). As can be seen from the plots, muon beam dump experiments exclude the region between $m_{\phi} \sim 1$ MeV and 30 MeV. On the other hand, the window between $m_{\phi} \sim 30 - 200$ MeV is still open. This window is projected to be explored by the proposed NA64 project at CERN \cite{Andreas:2013lya, Gninenko:2014pea}. The NA64 experiment is a fixed-target experiment that can run in the muon mode with a beam energy of 160 GeV and is designed for searching for missing energy $\gtrsim 50$ GeV. This experiment can help probe this open window. The projected region in the parameter space is shown by the dashed line in Fig. (\ref{fig:limits}).

Another proposed experiment is Fermilab's displaced decay search with a muon beam energy of 3 GeV \cite{Chapelain:2017syu}. However, the projected sensitivity of this experiment covers only a part of the projected sensitivity of the NA64 experiment, therefore I will not plot it here.

\subsection{Constraints from LEP Mono-$\gamma$ and $ee \rightarrow \tau\tau\gamma$ Searches}
\begin{figure}[!ht] 
  \centering
    \includegraphics[width=0.4\textwidth]{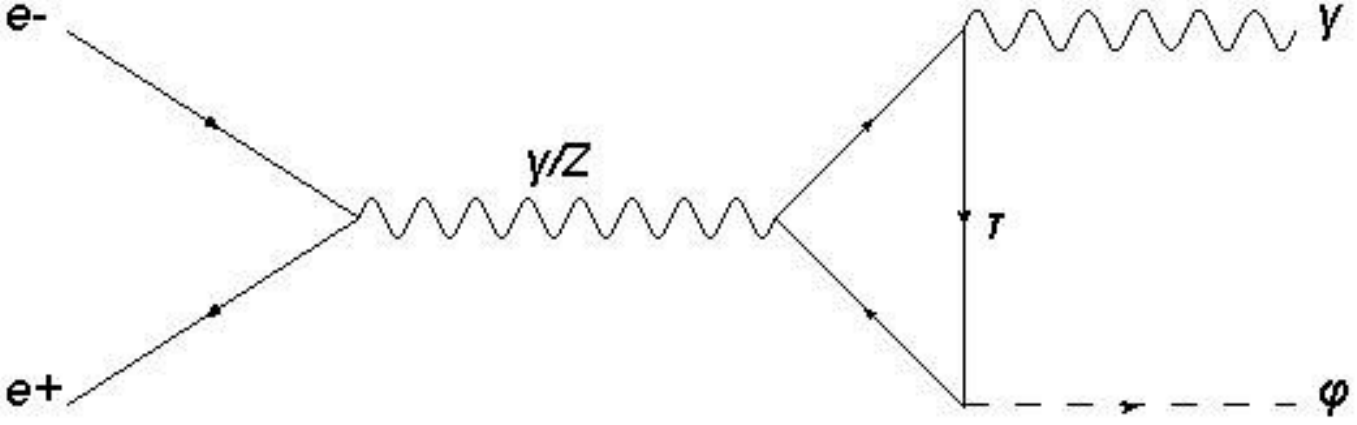}
      \caption{$e^{+}e^{-} \rightarrow \gamma\phi$ triangle diagram.}
      \label{fig:Triangle}
\end{figure}

The LEP mono-$\gamma$ searches were conducted to set limits on the number of neutrinos via studying the process $e^{-}e^{-} \rightarrow \nu \bar{\nu}\gamma$. The results can be used to set limits on the $m_{\phi} - g_{\phi ll}$ parameter space by considering the process $e^{+}e^{-} \rightarrow \gamma \phi$ for the range $m_{\phi} \leq 2m_{e}$. The tree-level process proceeds through the $t$ and $u$ channels, however, due to the smallness of the scalar's coupling to the electron, the constraints are weak. On the other hand, stronger constraints can be obtained through the triangle diagram shown in Fig. (\ref{fig:Triangle}) where the $\tau$ runs in the loop.

Similar to the case of the Higgs, we can write the effective Lagrangian as:

\begin{equation}\label{eq:HiggsEffectiveL}
\mathcal{L}_{eff} = c_{\gamma} \frac{\alpha}{\pi v_{\phi}} A_{\mu\nu}A^{\mu\nu} \phi + c_{\gamma Z} \frac{\alpha}{\pi v_{\phi}} A_{\mu\nu}Z^{\mu\nu}\phi,
\end{equation}
where $v_{\phi} = g_{\phi \tau\tau}^{-1}$. The effective couplings can be extracted from the results of the Higgs. For example, we can use the result of the Higgs decay to $\gamma\gamma$ where instead of the top running in the loop, we have the tau. This gives:
\begin{equation}\label{eq:cgamma}
c_{\gamma} = \frac{1}{8}|A^{\phi}_{\tau}(\tau_{\tau})|,
\end{equation}
where $\tau_{\tau} = \frac{4m_{\tau}^{2}}{m_{\phi}^{2}}$, and:
\begin{equation}\label{eq:A1}
A^{\phi}_{\tau}(\tau) = 2\tau \Big[1+ (1-\tau)f(\tau) \Big],
\end{equation}

\begin{equation}
         f(\tau)  =
        \left\{ \begin{array}{lll}
            \Big[\sin^{-1}\Big({\frac{1}{\sqrt{\tau}}\Big)} \Big]^{2} & , &  \text{$ \tau \geq 1$} \\
           -\frac{1}{4}\Big[\ln{\Big( \frac{1+\sqrt{1-\tau}}{1-\sqrt{1-\tau}}\Big)} -i \pi \Big]^{2} & , & \text{$\tau < 1$.} 
        \end{array} \right.
    \end{equation}

Similarly, we find $c_{\gamma Z}$ from the Higgs decay to $Z \gamma$. We obtain:
\begin{equation}\label{cgammaZ}
c_{\gamma Z} = \frac{1}{4\sin{\theta_{W}}} |B^{\phi}_{\tau}(\tau_{\tau}, \lambda_{\tau})|,
\end{equation}
where $\lambda_{\tau} = \frac{4m_{\tau}^{2}}{m_{Z}^{2}}$ and:
\begin{equation}\label{eqA2}
B_{\tau}^{\phi}(\tau,\lambda) = 2N_{c}^{\tau} \Bigg[ \frac{Q_{\tau} \big( I_{3}^{\tau} - 2Q_{\tau} s_{w}^{2}\big)}{c_{w}}  \Bigg] \Big[I_{1}(\tau,\lambda) - I_{2}(\tau,\lambda)\Big],
\end{equation}
\begin{multline}\label{eq:I1}
I_{1}(\tau,\lambda) = \frac{\tau \lambda}{2(\tau - \lambda)} +\frac{\tau^{2} \lambda^{2}}{2(\tau - \lambda)^{2}}\big[ f(\tau) - f(\lambda) \big] \\ +\frac{\tau^{2} \lambda}{(\tau -\lambda)^{2}} \big[g(\tau)-g(\lambda) \big],
\end{multline}
\begin{equation}\label{eqI2}
I_{2}(\tau,\lambda) = - \frac{\tau \lambda}{2(\tau - \lambda)} \big[ f(\tau) - f(\lambda)\big],
\end{equation}

\begin{equation}
         g(\tau)  =
        \left\{ \begin{array}{lll}
             \sqrt{\tau -1} \sin^{-1} \big({\frac{1}{\sqrt{\tau}}}\big) & , &  \text{$ \tau \geq 1$} \\
           \frac{\sqrt{\tau -1}}{2} \Big[\ln{\Big( \frac{1+\sqrt{1-\tau}}{1-\sqrt{1-\tau}}\Big)} -i \pi \Big] & , & \text{$\tau < 1.$} 
        \end{array} \right.
    \end{equation}
Here $Q_{\tau} =-1$, $N_{c}^{\tau} = 1$ and $I_{3}^{\tau} = -\frac{1}{2}$. Armed with this, we can find the cross-section of the triangle diagram in Fig. (\ref{fig:Triangle}):
\begin{equation}\label{eq:TriangleXsection}
\begin{split}
\sigma(e^{+}e^{-} \rightarrow \gamma \phi) = \Big(\frac{e^{2} g_{\phi ll}}{32\pi^{2} \sqrt{2\pi}c_{\text{w}}} \Big)^{2} \Big( \frac{1-m_{\phi}^{2}/s}{1-m_{Z}^{2}/s}\Big)^{2} \\
\times \Big[(1-4 s_{\text{w}}^{2}+ 8 s_{\text{w}}^{4} ) g^{2} c_{\gamma Z}^{2} + 32  c_{\text{w}}^{2}e^{2} c_{\gamma}^{2} (1-m_{Z}^{2}/s)^{2} \\
- 8 \hspace{0.5 mm} c_{\text{w}} \hspace{0.5 mm} g \hspace{0.5 mm}e\hspace{0.5 mm} c_{\gamma} c_{\gamma Z}(1-4s_{\text{w}}^{2})(1-m_{Z}^{2}/s)  \Big],
\end{split}
\end{equation}
where $s_{\text{w}}, c_{\text{w}}$ are the sine and cosine of the Weinberg angel respectively. The results from ALEPH \cite{Heister:2002ut, Barate:1998ci}, L3 \cite{Acciarri:1998vf, Acciarri:1997dq, Achard:2003tx}, OPAL \cite{Ackerstaff:1998ic} and DELPHI collaborations \cite{Abdallah:2003np} can be used in order set constraints on the scalar by requiring that at a $2\sigma$ level:
\begin{equation}\label{eq:LEPBounds}
\sigma(e^{+}e^{-} \rightarrow \gamma \phi) < 2 \delta_{\sigma}^{\text{Measured}},
\end{equation}
for $m_{\phi} \leq 2m_{e}$. Notice that for $m_{\phi} \leq 2m_{e}$, $\phi$ could decay to $\gamma\gamma$, however, for both benchmark points, the decay length is orders of magnitude larger than the dimensions of the detector, so $\phi$ appears as missing energy. On the other hand, for $m_{\phi} > 2m_{l}$, where $l = e$, $\mu$, $\tau$, $\phi$ can decay to a lepton pair, so we can use LEP $ee \rightarrow ll +\gamma$ searches for that region. The cross-section for $ee \rightarrow ll\gamma$ can be readily obtained by multiplying the cross-section in Eq. \ref{eq:TriangleXsection} by the appropriate branching fraction for the mass range.  The constraints are shown in brown in Fig. (\ref{fig:limits}). 

\subsection{\label{sec:Other}Constraints from Solar Emission, HB Stars and SN1987A}
Supernova 1987 (SN1987A), HB stars and solar emission constraints can impose stringent constraints on ALPs for masses $\lesssim 1$ GeV (see \cite {Jaeckel:2015jla, Cadamuro:2011fd} for instance). However, such constraints are only relevant if the ALP's dominant coupling is to photons. Since the coupling to photons in the type of models we are considering in this paper is favored to be much less than that to leptons (at least for the scalar case), those limits need to be revisited. 

If $\phi$ is produced in the sun or in HB stars, then it could affect the measured energy loss rate of the star when it streams out, in addition to affecting the star's evolution. Therefore one can obtain a bound on $m_{\phi}$ and $g_{\phi ll}$ by requiring that the amount of energy carried away by the scalar be less than the observed limits. 

In this model where the scalar's couplings to leptons are dominant, $\phi$ will be mainly produced in stars via its interaction with electrons. Assuming that the electrons are nonrelativistic and nondegenerate, which is a good assumption for the solar and  HB stars' mediums, the Compton-like scattering $e^{-} \gamma \rightarrow e^{-} \phi$ dominates over both Bremsstrahlung $e^{+}e^{-} \rightarrow e^{+}e^{-}\phi$ and electron-positron annihilation. For $m_{\phi} \ll m_{e}$, we can neglect the recoil energy of the electron and to a good approximation the energy loss rate per unit volume is given by \cite{Raffelt}:

\begin{equation}\label{eq:ELossRate}
Q = \frac{n_{e}}{\pi^{2}} \int_{m_{\phi}}^{\infty} d\omega \frac{\sigma(\omega) \omega^{3}}{e^{\omega/T}-1},
\end{equation}
where $\omega$ is the energy of the photon and $n_{e}$ is the number density of the electrons, which in terms of the electron fraction in the star $y_{e}$, the mass density of the star $\rho$, and the atomic mass unit $m_{u}$, is given by:
\begin{equation}\label{eq:eNumberDensity}
n_{e} = \frac{y_{e} \rho}{m_{u}}.
\end{equation}
 
\begin{figure}[!ht] 
  \centering
  \begin{minipage}[t][5cm][t]{0.4\textwidth}
    \includegraphics[width=\textwidth]{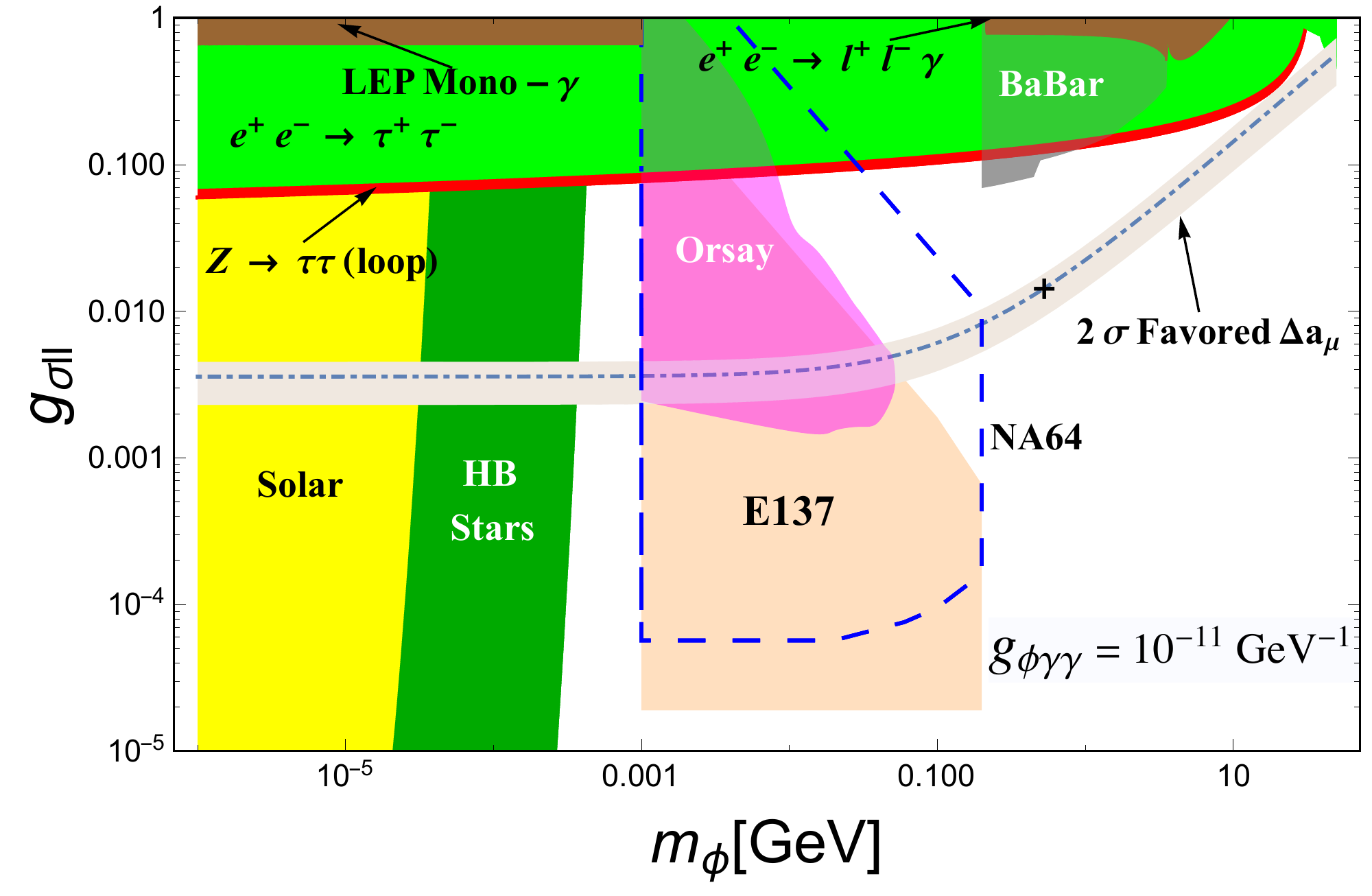}
  \end{minipage}
  \begin{minipage}[t][5cm][t]{0.4\textwidth}
    \includegraphics[width=\textwidth]{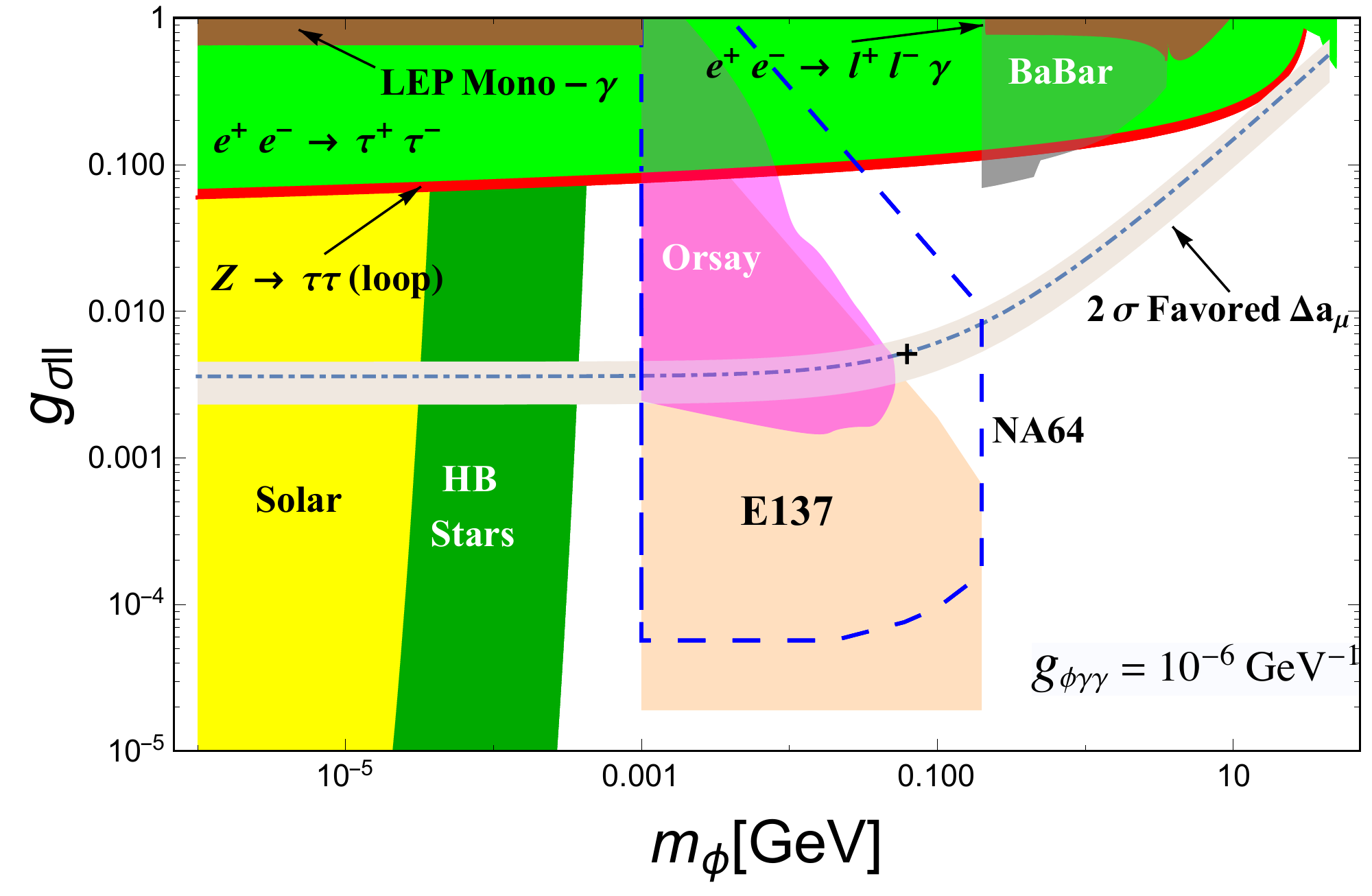}
  \end{minipage}
      \caption{The exclusion plot of the $m_{\phi} -g_{\phi ll}$ parameter space corresponding to the $g_{\phi\gamma\gamma} = 10^{-11}$ GeV$^{-1}$ (top) and $g_{\phi\gamma\gamma} = 10^{-6}$ GeV$^{-1}$ (bottom) benchmark points. The plot shows the $2\sigma$ allowed region for $\Delta a_{\mu(e)}$ (beige); the excluded region by the NLO correction to the leptonic $Z$ decay (red), the excluded region due to the NLO corrections to $ee \rightarrow \tau\tau$ (light green), Orsay (magenta), E137 (orange), LEP mono-$\gamma$ and $ee \rightarrow ll + \gamma$ searches (brown), solar emission (yellow) and HB star emission (dark green); and the projected region for NA64 (dashed). The plots also show the favored points that correspond to the minimum $\chi^{2}$ in the two scenarios.}
      \label{fig:limits}
\end{figure}

This calculation was done by Grifols \& Mass\'o in \cite{Grifols:1986fc} with $m_{\phi}$ neglected. Here, I will keep the mass of the scalar but assume it's less than $m_{e}$, and keep the assumption that the recoil energy of the electron is small. The cross-section is given by:
\begin{equation}\label{eq:ComptonXsection}
\sigma(e^{-} \gamma \rightarrow e^{-} \phi) = \frac{e^{2}g_{\phi ll}^{2}}{32 \pi} \frac{m_{e}^{2}}{\omega^{2}}F(x,y),
\end{equation}
where $x = \frac{m_{\phi}}{\omega}$, $y = \frac{m_{e}}{\omega}$ and $F(x,y)$ is some complicated function that I relegate to Appendix \ref{Appendix}. The bound on the solar (HB) emission rate is \cite{Raffelt}
\begin{equation}\label{eq:SolarHBBounds}
\dot{\epsilon} \leq 2(10) \hspace{2mm} \text{erg g$^{-1}$ s$^{-1}$}.
\end{equation}
Given this bound, we can use Eq. \ref{eq:ComptonXsection} to solve Eq. \ref{eq:ELossRate} numerically in order to find the excluded region in the parameter space. Here I set the average temperature to be $10^{7}(10^{8})$ K for the solar (HB) medium, the average density to be $10^{2}(10^{4})$ g cm$^{-2}$ for the solar (HB) medium and use $y_{e} = 0.5$ for both. The yellow region in Fig. (\ref{fig:limits}) shows the excluded part of the parameter space by the solar constraints, while the dark green region shows the excluded part by HB stars. As the plot shows, all masses below $\sim 350$ KeV are excluded. However, we must note that for $m_{\phi}$ close to the electron mass, the bound is less rigorous as our assumption of a small electron's recoil energy becomes less valid. However, I checked numerically that including the recoild energy of the final state enelctron does not significantly impact the bound. Also, notice that the HB constraints are much more stringent than the solar ones in spite of the weaker bound in Eq. \ref{eq:SolarHBBounds}. This is due to the higher temperature of HB stars compared with the sun, which yields a larger Boltzmann factor in Eq. \ref{eq:ELossRate}.

Of course, this analysis is valid only if $\phi$ streams freely out of the star, i.e. if it does not get trapped inside the medium of the star. In order to verify the validity of this assumption, we can calculate the mean free path of $\phi$ and compare it with the radius of the star. The mean free path is given by:
\begin{equation}\label{eq:MeanFreePath}
\lambda \sim \frac{1}{\sigma n_{e}} = \Big[ \frac{e^{2} g_{\phi ll}^{2}}{32\pi m_{u}} m_{e}^{2}\omega^{2} y_{e} \rho F(x,y) \Big]^{-1}.
\end{equation}

Assuming that the average photons energy is given by:
\begin{equation}\label{eq:GammaE}
\braket{\omega} = \frac{\pi^{4}}{30\zeta(3)}T \simeq 2.701T,
\end{equation}
one can easily check that for the entire mass range of interest, the mean free path is orders of magnitude larger than the radius of the sun or the typical radius of an HB star, thereby justifying the free streaming assumption.

A similar argument applies for scalars produced in supernovas. However, it was shown in \cite{Essig:2010gu} that for a scalar produced in the supernova core, the mean free path is given by:
\begin{eqnarray}\label{eq:SNmfp}
\lambda_{\text{mpf}} \sim 10 \hspace{1mm}\text{m} \Big( \frac{g_{\phi ll}^{-1}}{10^{6} \hspace{1mm} \text{GeV}} \Big)^{2},
\end{eqnarray}
which means that for $g_{\phi ll}$ larger than $10^{-6}$ GeV$^{-1}$, the scalar gets trapped in the core and never streams out. Thus, SN1987A does not costrain the favored region and therefore we ignore it.

\subsection{\label{sec:Discussion}Discussion}
We have shown that the constraints from BaBar, beam dump experiments, the NLO correction to the $Z$ decay, LEP mono-$\gamma$ searches, $ee\rightarrow \tau\tau(\gamma)$ searches, and solar and HB star constraint, exclude a significant part of the parameter space. Fig. (\ref{fig:limits}) shows that all masses below $\sim 350$ KeV are excluded by solar and HB constraints; beam dump experiments exclude the region above 1 MeV up to $\sim 30$ MeV, while masses above $\sim$ 10 GeV are in significant tension with the NLO corrections to the $Z$ decay to $\tau\tau$ and the NLO corrections to $ee \rightarrow \tau\tau$, although this region is not entirely excluded. This leaves a large open window between $\sim 30$ MeV and $\sim 10$ GeV that is most favored to be explored.

Part of this region is projected to be explored by the proposed NA64 experiment, which is projected to cover the mass range from 1 MeV and up to the di-muon mass. The first benchmark point corresponding to $g_{\phi\gamma\gamma} = 10^{-6}$ GeV$^{-1}$ lies in this region. On the other hand, the second benchmark point lies outside this region but could be explored by the Belle II experiment \cite{Abe:2010gxa}. The Belle II experiment has recently started collecting data and it is expected that by 2025, it would have reached a total integrated luminosity of $50 \hspace {1mm} \text{ab}^{-1}$. This projects it to be more sensitive than the BaBar experiment, and therefore might help explore more of the parameter space above the dimuon threshold which is not covered by NA64 and up to $\sim 5$ GeV. This leaves two windows, one above $\sim 5$ and the other is between $\sim 350$ KeV and 1 MeV. 

The International Linear Collider (ILC) \cite{Behnke:2013xla} (if built) might help explore both regions. For the former, the ILC's ultra-precision measurement of the $Z$ production and subsequent decay can help improve the bounds from the $Z$ decay by lowering the measured uncertainties, while for the latter region, the $e^{+}e^{-} \rightarrow h+ \cancel{E}_{T}$ channel can help investigate the hypothetical process $e^{+}e^{-} \rightarrow h+ \phi$ .

A final point to mention is that due to the assumption made in Eq. \ref{eq:YukawaCouplings}, the dominant coupling will be to the tau lepton, and therefore collider and beam dump experiments limits are more stringent. If we assume a different type of coupling to leptons, say by assuming a suppressed coupling to the tau compared to the muon, then the constraints will be alleviated, and more of the parameter space will open. We will very briefly discuss one model where this can be achieved in the next section.

\section{\label{sec:Radion}A Radion Solution for the $g-2$ Anomaly?}
In the Randall-Sundrum (RS) model \cite{Randall:1999ee}, the radion is the scalar field that parametrizes the fluctuations of the extra dimension around its potential minimum. The radion could pose an interesting possibility for solving the $g-2$ anomaly due to its unique couplings to matter. More specifically, the radion's coupling to matter is highly model-dependent and varies according to the localization of the matter fields on either of the branes or in the bulk. The coupling to brane-localized matter is given by \cite{Csaki:2000zn}:

\begin{equation}\label{eq:radionCoupling}
\frac{\phi(x)}{\Lambda_{UV,IR}} \text{Tr} T_{\mu\nu},
\end{equation}
where $\phi (x)$ is the 4D radion field, $\Lambda_{UV,IR}$ is the radion constant on the UV and IR branes respectively and $T_{\mu\nu}$ is the stress-energy tensor. Since the UV scale is typically many orders of magnitude larger than the IR scale, it is possible to suppress the coupling to the tau lepton compared to the muon by assuming that the former is localized on the UV brane, while assuming that the latter is localized on the IR brane. This way, one could alleviate all of the constraints (except for beam dump experiments) as the couplings will be rescaled by $m_{\mu}/m_{\tau}$.

Although the typical mass of the radion is comparable to the Electroweak (EW) scale ($\sim 100$ GeV), much lighter masses can be achieved through the Contino-Pomarol-Rattzzi (CPR) mechanism \cite{CPR} as was demonstrated in \cite{Abu-Ajamieh:2017khi, Abu-Ajamieh:2018brk}.

Another interesting aspect of a radion solution is that the radion could couple to nucleons and pions through quarks and gluons \cite{Abu-Ajamieh:2017khi}, which presents additional experimental probes. Focusing on the coupling to pions, we can write the effective Lagrangian as:

\begin{equation}\label{eq:PionEffective}
\mathcal{L_{\phi\pi\pi}} = g_{\phi\pi\pi}m_{\pi}^{2} \phi\pi^{+}\pi^{-},
\end{equation}
where $ g_{\phi\pi\pi}$ is the effective radion coupling to pions with dimension $(mass)^{-1}$.  If the radion is heavy enough, it could decay to $\pi^{+}\pi^{-}$:

\begin{equation}\label{eq:DecayToPions}
\Gamma(\phi \rightarrow \pi^{+}\pi^{-}) = \frac{g_{\phi\pi\pi}^{2}}{16 \pi} \frac{m_{\pi}^{4}}{m_{\phi}} \Bigg( 1-\frac{4m_{\pi}^{2}}{m_{\phi}^{2}} \Bigg)^{\frac{1}{2}}.
\end{equation}

The decay width in Eq. \ref{eq:DecayToPions} could be small for typical values of $g_{\phi\pi\pi}$ but could still be measurable. For instance, the decay width for a 400 MeV radion with $g_{\phi\pi\pi} = 0.3$ GeV$^{-1}$ would be $\simeq 1.2$ KeV. 

For lighter masses, searches for the rare pion decay $\pi^{-} \rightarrow \mu^{-} \bar{\nu}_{\mu} \phi$ could provide an interesting search option. If we assume that $g_{\phi\pi\pi}$ dominates over $g_{\phi ll}$, then the branching fraction of this hypothetical decay would be given by:

\begin{equation}\label{eq:BrPion}
\text{Br}(\pi \rightarrow \mu \nu \phi) \simeq 1.35 \times 10^{-2} \Big(\frac{g_{\phi\pi\pi}}{\text{GeV}^{-1}} \Big)^{2} \%.
\end{equation}

For instance, $g_{\phi\pi\pi} \sim 0.03$ GeV$^{-1}$ would yield a branching fraction comparable to the observed rare decay $\pi^{+} \rightarrow e^{+}e^{-}e^{+}\nu_{e}$.
\section{\label{sec:Conclusions}Conclusions}
The $g-2$ anomaly remains one of the best ways to search for physics BSM. In this paper, I investigated a class of models with a scalar/pseudoscalar that has a coupling to leptons proportional to the lepton's mass. 

We saw in this paper that for the case of a pseudoscalar solution, there is no region in the parameter space that could simultaneously solve both the electron and the muon anomalies with Yukawa couplings of the form in Eq. \ref{eq:YukawaCouplings}. However, if no assumption is made regarding the form of the Yukawas, it is possible to have a pseudoscalar solution for both of the anomalies. Nevertheless, this solution is not very attractive since it would require tuning the scalar's coupling to photons to be somewhat large in order for both of the Yukawa couplings to have the same sign. Such a large coupling to photons would be disfavored by cosmological observations.

On the other hand, a scalar can simultaneously provide a solution for both anomalies while having the required form of the Yukawa couplings. In such a case, we demonstrated that such a solution favors smaller couplings to photons $\lesssim 10^{-6}$ GeV$^{-1}$, and we established the corresponding favored region in the $m_{\phi} - g_{\phi ll}$ parameter space that corresponds to two representative benchmark points. We investigated the experimental constraints from the BaBar experiment, beam dump experiments, the NLO corrections to the $Z$ decay and to $ee \rightarrow \tau\tau$, LEP mono-$\gamma$ and $ee \rightarrow ll+\gamma$ searches, and from solar and HB emission bounds, and we saw that a significant part of the parameter space is excluded. In particular, we found only two open windows for $m_{\phi}$ between $\sim 350$ KeV and 1 MeV, and between $\sim 30$ MeV and $\sim 50$ GeV, with the region above $\sim 10$ GeV being in significant tension with experiment.

The tools used in this paper can be used to constrain other solutions to the $g-2$ anomaly, such as solutions that adopt the $Z'$ or the dark photon to explain it. I expect that the limits on these solutions would not be too different from the scalar case for the same range of masses and couplings, they are nonetheless worthwhile investigating. 

Future experiments, such as NA64, Belle II, and the ILC can help explore significant regions of the parameter space, and one hopes that in the near future, enough data would be collected to shed more light on the remaining open windows, and thus help explore the viability of this solution.

\section*{Acknowledgments}
I would like to thank John Terning for his valuable insight. I would also like to thank John Conway, Lloyd Knox and Max Chertok for answering my questions.
\\
\\
\appendix

\section{Explicit Expression for $F(x,y)$} \label{Appendix}
\begin{multline}
F(x,y) = \frac{2y}{(1+2y+2y^{2})^{4}(1+4y^{4}+4 x^{2}y^{2})}\\ \times \Bigg[(1+2y+2y^{2})(1+2y+2y^{2}+4y^{3})^{2}(1+4y^{4}+4x^{2}y^{2})\\
\times \log{\Big(  \frac{1+2y^{2}+2y \sqrt{1-x^{2}}}{1+2y^{2}-2y \sqrt{1-x^{2}}} \Big)} -4y\sqrt{1-x^{2}}\Big(16y^{8}(2x^{2}+9) +\\
16y^{7}(4x^{2}+7) +16y^{6}(5x^{2}+4) + 40 y^{5}(2x^{2}+1) +24 y^{4}(2x^{2}+1) +\\
4y^{3}(6x^{2}+5)+4y^{2}(x^{2}+3)+64y^{10}+128y^{9}+6y+1 \Big) \Bigg].
\end{multline}
\\

\textbf{Added note:} After finalizing this paper, a new results \cite{Aoyama:2019ryr} was published claiming a $2.4 \sigma$ discrepancy in the measured electron anomaly:
\begin{equation}
\Delta a_{e} = - 88 \pm 36 \times 10^{-14}.
\end{equation}

If this yet uncorroborated result is true, then our results cannot explain both anomalies simultaneously at the $2\sigma$ level, although the results remain valid at the $3\sigma$ level. This is because the electron anomaly will be negative at the $2\sigma$ level and a novel explanation would be needed to explain the sign discrepancy between the two anomalies. In this case, the electron anomaly in this work would serve more as a constraint on the allowed region of muon anomaly, and the result in Figure 9 would be relevant for the muon anomaly only. Either way, a corroborating result is still needed to confirm the value and the sign of the electron anomaly.

\end{document}